\documentclass[showpacs,prl,onecolumn,aps,superscriptaddress,preprintnumbers,nofootinbib]{revtex4}
\usepackage[T1]{fontenc}
\usepackage[latin9]{inputenc}
\usepackage{amsmath,amssymb,bigints}
\usepackage{epsfig}
\usepackage{graphicx}
\usepackage{amsmath}
\usepackage{amsfonts}
\def\slashchar#1{\setbox0=\hbox{$#1$}     		% set a box for #1
   \dimen0=\wd0                                 	% and get its size
   \setbox1=\hbox{/} \dimen1=\wd1               	% get size of /
   \ifdim\dimen0>\dimen1                        	% #1 is bigger
      \rlap{\hbox to \dimen0{\hfil/\hfil}}      	% so center / in box
      #1                                        	% and print #1
   \else                                        	% / is bigger
      \rlap{\hbox to \dimen1{\hfil$#1$\hfil}}   	% so center #1
      /                                         	% and print /
   \fi}

\renewcommand{\vec}{\boldsymbol}
\newcommand{\beq}{\begin{equation}}
\newcommand{\eeq}{\end{equation}}
\newcommand{\bea}{\begin{eqnarray}}
\newcommand{\eea}{\end{eqnarray}}
\newcommand{\baa}{\begin{array}}
\newcommand{\eaa}{\end{array}}

\def\eq#1{{Eq.~(\ref{#1})}}
\def\fig#1{{Fig.~\ref{#1}}}

\newcommand{\bas}{\bar{\alpha}_S}
\newcommand{\as}{\alpha_S}

\newcommand{\nn}{\nonumber}

\newcommand{\h}{\frac{1}{2}}

\newcommand{\Lb}{\left(}
\newcommand{\Rb}{\right)}

\renewcommand{\vec}[1]{\boldsymbol{#1}}

\begin{document}
%%%%%%%%%%%%%%%%%%%%%%%%%%%%%%%%%%%%%%%%%%%%%%%%%%%%%%%%%%%%%%%%%%%%%%%
\title{ Thermal radiation  and inclusive production in  the CGC/saturation
 approach at high energies.}
\author{ E.~ Gotsman,}
\email{gotsman@post.tau.ac.il}
\affiliation{Department of Particle Physics, School of Physics and Astronomy,
Raymond and Beverly Sackler
 Faculty of Exact Science, Tel Aviv University, Tel Aviv, 69978, Israel}
 \author{ E.~ Levin}
 \email{leving@post.tau.ac.il, eugeny.levin@usm.cl} \affiliation{Department of Particle Physics, School of Physics and Astronomy,
Raymond and Beverly Sackler
 Faculty of Exact Science, Tel Aviv University, Tel Aviv, 69978, Israel} 
 \affiliation{Departemento de F\'isica, Universidad T\'ecnica Federico Santa Mar\'ia, and Centro Cient\'ifico-\\
Tecnol\'ogico de Valpara\'iso, Avda. Espana 1680, Casilla 110-V, Valpara\'iso, Chile} 

\date{\today}


\begin{abstract}
In this paper, we discuss the inclusive production of hadrons in the
 framework of the  CGC/saturation approach. We argue, that  gluon jet  
inclusive production stems from the vicinity of the saturation momentum,
 even for small values of the transverse momenta $p_T$. Since in this  
region, we theoretically,  know the scattering amplitude,  
we claim that we
 can provide  reliable estimates for this process. We demonstrate,
 that in a widely  accepted  model for  confinement, we require  a 
thermal radiation term to describe the experimental data. 
 In this model  the parton (quark or gluon)  with the transverse
 momenta of the order of $Q_s$ decays into  hadrons with  the
 given fragmentation functions, and the production of the hadron with
 small transverse momenta is suppressed by the mass of the gluon  jet. 
 In addition we show that other approaches
 for  confinement,  also describe the data, without the need for 
 thermal emission.

\end{abstract}

\preprint{TAUP - 3030/18}

\maketitle

\tableofcontents

\flushbottom

%%%%%%%%%%%%%%%%%%%%%%%%%%%%%%%%%%%%%%%%%
\section{Introduction}

%%%%%%%%%%%%%%%%%%%%%%%%%%%%%%%%%%%%%%%%%%%%%%%%%%%%%
In this paper we discuss the dynamics of  generating
multi-hadron 
 processes at high energy in the framework of the Color Glass
 Condensate(CGC)/saturation approach (see Ref.\cite{KOLEB} for the review).
  These processes occur at long distances  and  therefore to treat 
them theoretically,
  we need to develop  a non-perturbative QCD approach. This is a very
 difficult and challenging problem, which is far from being solved.  The
 CGC/saturation approach, being an effective QCD theory at high energies,
 states that the new phase of QCD: the dense system of partons (gluons and
 quarks) is produced in   collisions  with a new characteristic
 scale: saturation momentum $Q_s(W)$,  which increases as a 
 function of energy $W$\cite{GLR,MUQI,MV}. However, the  transition
 from this system of partons to the measured state of hadrons is still
 an unsolved problem.
At the moment, we need to use 
  pure phenomenological input for 
the long distance non-perturbative physics, due to our lack of 
 theoretical understanding of the confinement of quarks and gluons. In
 particular,  we wish to use phenomenological fragmentation
 functions.  Hence, our model for confinement is that the parton
 (quark or gluon)  with the transverse momenta of the order of $Q_s$
 decays into  hadrons with  the given fragmentation functions. The
 experimental data confirm this model of hadronization, which is the
 foundation of all Monte Carlo simulation programs, and  leads to
 descriptions of the 
 transverse momenta distribution of the hadrons at the LHC energies.
  As an example, we refer to Ref.\cite{ALICE13}, which shows that the
 next-to-leading order QCD calculations with formation of the hadrons
 in accord with the fragmentation functions (\cite{NLOFIT,NLOFIT1}),  is able
 to describe the transverse momentum spectra for the LHC range of
 energies.  However, such a  description is only successful for large 
values
 of $p_T\, >\, 3\, GeV$\cite{NLOFIT} 
 or $p_T\, >\, 5 GeV$\cite{NLOFIT1}, while we assume that one can use 
these
 fragmentation functions in the region of small $p_T$ as well.
 In a sense, at present, this model is the best that we 
can  propose
  to describe   multi - hadron production. 
 
It turns out\cite{FIT0,FIT1,FIT2,FIT3,BAKH,FPV}  that the  experimental
 data \cite{ALICE13,ALICE2,ATLAS1,ATLAS2,CMS1,CMS2} at high energy, can
 be describes as the sum of two terms:
\beq \label{SUM}
\frac{d \sigma}{dy d^2 p_T}\,\,=\,\,\underbrace{A_{\rm therm} e^{- \frac{m_T}{T_{\rm th}}}}_{\mbox{ thermal radiation}}\,\,\,\,+\,\,\,\,\underbrace{A_{\rm hard}\frac{1}{\Lb 1 + \frac{m^2_T}{T^2_{\rm h}\,n}\Rb^n}}_{\mbox{hard emission}}
\eeq
with
\beq \label{T} 
 T_{\rm th}\,\,=\,\,0.098 \Lb \sqrt{\frac{s}{s_0}}\Rb^{0.06} \,{\rm GeV};~~~~~~~~~~
  T_{\rm h}\,\,=\,\,0.409 \Lb \sqrt{\frac{s}{s_0}}\Rb^{0.06} \,{\rm GeV}; 
  \eeq
 
 We believe, on a qualitative level,
 that these  two terms have  a natural explanation in the
 CGC/saturation approach. The
 second term has  a  power- like decrease ($\propto  1/p^{2n}_T$)
 at large $p_T$,  as it should be in perturbative QCD. 
 The appearance
 of a thermal term in  a high energy proton-proton collision is a
 remarkable feature of the interaction, since the number of the
 secondary interactions in proton-proton collisions is rather
 low, and cannot provide the thermalization due to the interaction 
in the final state.  Therefore, the appearance of the first term
 in \eq{SUM} is not related to the equilibrium of the produced system
 of partons, and we used the word `thermal' just to characterize the
 form of $p_T$ dependence of this term.
  The origin of  thermal radiation in the
 framework of the CGC approach was clarified a decade ago
 \cite{KHTU,DUHA,CKS,KLT} and, recently, the new idea that
 the quantum entanglement is at the origin of the parton densities
 has been added to these arguments\cite{KHLE}. The resulting picture
 is  presented nicely in Ref.\cite{BAKH}, to which we refer our
 readers. The brief sketch below, is  intended  to indicate
 the main ideas that  originate in the CGC approach.
 
 For proton-proton scattering in the lab. frame, the parton 
configuration in QCD is formed long before the interaction at
 distances $1/(m x)$, where $m$ - denotes the proton mass, and $x$ 
 the fraction of longitudinal momentum carried by parton which
 interacts with the target. However, before the collision, the wave
 function of this partonic fluctuaction is the eigenfunction of
 the Hamiltonian  and, therefore, the system has  zero entropy. 
The interaction with the target of  size $R$ destroys the coherence
 of the parton wave
function of the projectile. The typical time, which is needed for this,
 is of the order of
$\Delta t \propto R$, and is much smaller than the lifetime of all 
faster partons in the fluctuation. Hence, this interaction  can be
 viewed as a rapid quench of the
entangled partonic state\cite{KHLE} with substantial  entanglement
  entropy.  After this rapid quench, the interaction of the gluons 
 change the Hamiltonian.  In the CGC approach, 
all partons with rapidity larger than that of a particular gluon $y_i$, 
  live longer than this parton. They can be considered as the source
 of the classical
field that emits this gluon.  It was shown that  after the quench,
 the fast gluons create the  longitudinal chromo-electrical 
background field, which leads to the thermal distribution of the
 produced gluons. 
 
 The temperature of this distribution is intimately related to
 the saturation momentum, which provides  the only dimensional 
scale in the colour glass condensate. It determines both the strength
 of the longitudinal fields and the  ultraviolet cutoff on the
 quantum modes, resolved by the collision.  It turns out
 \cite{KHTU,KLT} that
 \beq \label{THTEM}
  T_{\rm th}\,\,=\,\,{\rm c} \frac{Q_s}{2 \pi}
  \eeq
  with the semi-classical estimates 
  \cite{KLT} for the constant $c = 1.2$. The saturation scale $Q_s$ depends
 on $x$ and the impact parameters ($b$) of the reaction, and has the
 following form\cite{GLR,MUQI,MV}:
  \beq 
  \label{QS}
  Q_s\Lb x, b\Rb\,\,=\,\,Q_0\Lb b \Rb \Lb \frac{1}{x}\Rb^{\lambda/2}\,\,=\,\,Q_s\Lb x\Rb \,S\Lb b \Rb
  \eeq
  The value of $\lambda $ can be calculated theoretically and measured
 experimentally. The leading order QCD evaluation leads to
 $\lambda = 4.9\bas$,where $\bas$ denotes the running QCD coupling.
 Plugging in the reasonable estimate for $\bas\Lb Q_s\Rb \approx 0.2$,
 one can see that $\lambda$ turns out to be large, about 0.8-1. The
 phenomenological description of the hard processes both for nucleus
 interactions\cite{DKLN} and DIS( see Ref.\cite{RESH} and references
 therein), give the value of $\lambda = 0.2-0.24$. We  therefore see, 
 that in the CGC approach we expect that $$T_{\rm th} \propto\,\,T_{\rm h} \propto\,\,Q_s \propto \Lb \sqrt{\frac{s}{s_0}}\Rb^{\lambda/2} \sim \Lb \sqrt{\frac{s}{s_0}}
\Rb^{0.1 - 0.112},  $$ \eq{SUM} and \eq{T} show that both temperatures 
have dependence on energy in accord with the CGC result but, on the other hand,
   this dependence  contradicts the CGC prediction  that this dependence
 should be proportional to the saturation scale. 
Especially,  the second term in \eq{SUM}, which
 corresponds to the 
  contribution of the hard processes  looks strange. Indeed, the 
above interpretation
 of the thermal radiation cannot be considered as the conventional one, 
 the CGC approach to the hard processes has been confirmed both
 theoretically and experimentally,  and  we know that the typical
 scale in these processes,  is the saturation momentum. The second
 remark is related to the value of the hard contribution. In the CGC
 approach  it should be calculated  theoretically,  and  not  be 
determined from 
 a  fitting procedure.
  
The goal of the paper is to re-visit  inclusive production in the 
 CGC/saturation approach for a  more thorough 
consideration,  and to show that the thermal term with the temperature
 given by \eq{THTEM}, is needed  for describing  the experimental
 data at high energies. It should be noted that in the first
 attempt\cite{LERE} to compare the CGC prediction with the
 experiment at $W\,=\,7 \,TeV$,  the thermal term was not required.

The paper is organized as follows. In the next section we discuss
 the general procedure for the calculation of the gluon inclusive
 production in CGC/saturation approach. In section III, we consider
 the evolution equation for the theory with a  simplified BFKL kernel. We
 show that the solution to this equation confirms our key  idea,
 that the main contribution to the inclusive production stems
 from the kinematic region in the vicinity of the saturation scale.
 Since theoretically we know the scattering amplitude in this region, 
 we demonstrate that  we are able to provide  reliable estimates
 for this process. In section IV we develop the saturation model which
 we need to use due to the long standing unsolved problem i.e. the 
behaviour
 of the scattering amplitudes at large impact parameter. In section V
 we compare our estimates with the experimental data, and demonstrate
 that within our model for confinement: the fragmentation function for
 the gluon jets,   needs to have a thermal radiation term with
 temperature, which is proportional to the saturation scale $Q_s$. In
 the Conclusions we summarize our results.

%%%%%%%%%%%%%%%%%%%%%%%%%%%%%%%%%%%%%%%%%%%%%%%%%%%%
\section{Inclusive production in CGC/saturation approach: generalities}

%%%%%%%%%%%%%%%%%%%%%%%%%%%%%%%%%%%%%%%%%%%%%%%%%%%%
The formula for  the gluon jet production in proton-proton collisions
 has the following general form (see Ref.\cite{KTINC} for the proof):

\bea \label{MF}
\frac{d \sigma_G}{d y \,d^2 p_{T}}\,\,& & \frac{2C_F}{\alpha_s (2\pi)^4}\,\frac{1}{p^2_T} \,\int d^2  r\,e^{i \vec{p}_T\cdot \vec{r}}\,\,\nabla^2_T\,\int d^2 b N_G\Lb y_1 = \ln(1/x_1); r, b \Rb\,\,\int d^2 b' \nabla^2_T\,N_G\Lb y_2 = \ln(1/x_2); r,  b' \Rb. 
\eea
where $N_G\Lb y_1 = \ln(1/x_1); r, b \Rb$ can be found from the amplitude
 of the dipole-proton scattering 
$N\Lb y_i = \ln(1/x_i); r_; b \Rb$ :
\beq \label{NG}
N_G\Lb y_i = \ln(1/x_i); r,  b \Rb\,\,=\,\,2 \,N\Lb y_i = \ln(1/x_i); r, b \Rb\,\,\,-\,\,\,N^2\Lb y_i = \ln(1/x_i); r,  b \Rb
\eeq
where $r$ denotes the size of the dipole, $b$  it's impact parameter 
and 
\beq \label{X}
x_1 \,\,=\,\,\frac{p_T}{W} \,e^{y};\,\,\,\,\,x_2 \,\,=\,\,\frac{p_T}{W}
 \,e^{-y};\eeq

where $y$ denotes the rapidity of the produced gluon in c.m.f. and $W$ the
 c.m.s. energy of the collision.  In this paper we consider the gluon
 production at $y = 0$.
$C_F = (N^2_c - 1)/2 N_c$ and $  \bas \,= \,\as N_c/\pi$ with the number
 of colours equals $N_c$. $\as$ denotes  the running QCD coupling, 
$\nabla^2_T$ 
  the Laplace operator with respect to $r$,  it is equal to 
$\nabla^2_T\,=\,\frac{1}{r} \frac{d}{d r} \Lb r \frac{d}{d r}\Rb$.

At high energies and sufficiently small values of $p_T$, the dipole 
amplitudes
 are in the saturation region, where the parton densities are large  and 
the
 dipole scattering amplitude displays  geometric scaling behaviour, being
 a function of  only one variable:  $ \tau \,=\,r \,Q_s\Lb W,b\Rb$.
 Introducing a new variable $
z\Lb r,b, x\Rb\,\,=\,\,\ln\Lb \tau^2\Rb$, we can re-write \eq{MF}
at $y=0$ in the
 form 
\bea \label{MF1}
&&\frac{d \sigma_G}{d y \,d^2 p_{T}}\,\,= \nn\\
&& \frac{2C_F}{\alpha_s (\pi)^4}\,\frac{1}{p^2_T} \,\int d^2  r\,e^{i \vec{p}_T\cdot \vec{r}}\,\,\,\int d^2 b\, Q^2_s\Lb x_1,b\Rb e^{- z\Lb r, b,x_1\Rb} \frac{d^2 N_G\Lb z(r, b, x_1))\Rb}{d z^2}\ \,\,\int d^2 b'\, Q^2_s\Lb x_1,b\Rb e^{- z\Lb r, b',x_2\Rb} \frac{d^2 N_G\Lb z(r, b', x_2)\Rb)}{d z^2}\nn\\
 &&= \frac{2C_F}{\alpha_s (\pi)^3}\,\frac{Q^2_s\Lb x \Rb}{p^2_T} \,\int d z \, e^{ - z\Lb r,b=0,x\Rb}\,J_0\Lb \frac{p}{Q(x)} e^{z}\Rb\,\,\,\Bigg(\int d^2 b\, S\Lb b \Rb  \frac{d^2 N_G\Lb z(r, b, x_1))\Rb}{d z^2}\Bigg)^2\nn\\
 &&\,\,\equiv\,\,\frac{2C_F}{\alpha_s (\pi)^3}\,\frac{1}{\tilde{p}^2_T} \int d z\,\,J_0\Lb \tilde{p}_T \,e^z\Rb\,\, I\Lb z\Rb\,\,=\,\,\,\,\frac{2C_F}{\alpha_s (\pi)^3}\,\frac{1}{\tilde{p}^2_T}\,\,{\cal I}\Lb \tilde{p}\Rb
  \eea
 In \eq{MF1} we have taken $x = x_1=x_2$ since $y = 0$, and introduce a 
new
 variable $\tilde{p}_T\,=\,p_T/Q_s\Lb x \Rb$.
 
  To calculate the hadron distributions, 
we need to take into account the  decay of the jet into hadrons.
  The formula has the form
 \beq \label{XSPION}
 \frac{d \sigma^{\pi}}{d y \,d^2 p_{T}} \,\,=\,\,\int^1_0  d x_G\, \frac{d \sigma_G}{d y \,d^2 p_{T}} \Lb \frac{p_T}{x_\pi}\Rb\,D^\pi_G\Lb x_\pi\Rb
 \eeq
 We take the fragmentation function $D^\pi_G$  from Ref.\cite{FRFU} 
 which has
 the form
 \beq \label{FRF}
 D^\pi_G\Lb x^\pi\Rb\,\,=\,\,2.17 z^{\alpha} (1-z)^{\beta} \left(20 (1-z)^{\gamma1}+1\right); 
 \eeq 
 
 with $\alpha = 0.899$,   $\beta= 1.57$ and $\gamma=4.91$.
 
 We note that \eq{MF1} as well as \eq{XSPION} leads to a cross
 section which is proportional to $1/p^2_T$. This behaviour results in a
 logarithmic divergency of the integral over $p_T$, or in other words
 gives an infinite number of produced pions at fixed rapidity.This 
divergency  also indicates that we need to reformulate our assumptions 
about   confinement, since using the fragmentation functions does not
 suppress the divergency at low $p_T$.
  We  believe that the reason
for this
 divergency, is the fact
   that we neglected the mass of the
 jet of hadrons that stem from the decay of the gluon.
  The simple
 estimates \cite{KHLEKLN} give  for a gluon with the value of the
 transverse momentum $p_T$,  the mass of the jet $m^2_{\rm jet}\,=\,2 
p_T\, m_{\rm eff}$, where $
  m_{\rm eff} = \sqrt{m^2 + k_T^2 + k^2_L} - k_L$, $m$ is the mass 
of
 the lightest hadron in the jet, $k_T$ is it's transverse momentum 
and $k_L
 \approx \,k_T$ is the longitudinal momentum of this hadron.
 Since  most pions stem from the decay of $\rho$-resonances we expect
 that $ m_{\rm eff} \approx m_\rho$.
 
 As we have mentioned  in the introduction,   our model for 
 confinement is that  of  the CGC approach, the typical momentum for the
 produced gluon is the saturation momentum. Hence,  most hadrons are
 created  in the jets with the mass $m^2_{\rm jet} \,=\,2 Q_s\,m_{\rm eff}$.
 However, for rare gluons with $p_T \gg  Q_s$ we still have $m^2_{\rm jet}
 \,=\,2p_T\,m_{\rm eff}$ . For  numerical  estimates we
 use $m^2_{\rm jet}\,=\,2 \Lb Q_s \Theta(  Q_s  -  p_T) + p_T 
\Theta( p_T  -  Q_s) \Rb m_{\rm eff}$ which has these two limits.
 $\Theta(x)$ denote the step function. Using the same idea we replace 
\eq{X} by
 
 \beq \label{XF}
 x \,\,=\,\,\frac{p_T}{W} \,\,=\,\,\frac{ Q_s \Theta(  Q_s  -  p_T)
 + p_T \Theta( p_T  -  Q_s) }{W}. 
 \eeq
 
 Hence, we finally deal with the following model for the confinement
 ( hadronization):  the decay of the gluon jet with the effective mass
 that we have discussed above, and with  fragmentation functions of
 \eq{FRF} at all  values of the transverse momenta.
 
%%%%%%%%%%%%%%%%%%%%%%%%%%%%%%%%%%%%%%%%%%%%%%%%%%%%
\section{Non-linear evolution for the  leading twist BFKL kernel }

\subsection{Equations}
%%%%%%%%%%%%%%%%%%%%%%%%%%%%%%%%%%%%%%%%%%%%%%%%%%%%

The dipole scattering amplitude is the solution of the Balitsky-Kovchegov
 \cite{BK} non-linear equation, which has the following form:

  \bea \label{BK}
\displaystyle{\frac{\partial N\Lb Y; \vec{x}_{01}, \vec{b}
 \Rb}{\partial Y}}\,&=&\,\displaystyle{\frac{\bas}{2\,\pi}\int d^2 
 \mathbf{x_{2}}\,K\Lb \vec{x}_{01};\vec{x}_{02},\vec{x}_{12}\Rb\Bigg\{
  N \Lb Y; \vec{x}_{02},\vec{b} - \h \vec{x}_{12}\Rb +  N\Lb Y; 
\vec{x}_{12},\vec{b} - \h \vec{x}_{02}\Rb\,-\,N\Lb Y; \vec{x}_{01
},\vec{b} \Rb}\,\nn\\
&-&\,\displaystyle{ N\Lb Y; \vec{x}_{02},\vec{b} - \h \vec{x}_{12}\Rb\,
 N\Lb Y ;\vec{x}_{12},\vec{b} - \h \vec{x}_{02}\Rb\Bigg\}}
~~~~~~\mbox{where}~~~~ \displaystyle{K\Lb \vec{x}_{01};\vec{x}_{02},
\vec{x}_{12}\Rb}\,\,=\,\,  \displaystyle{\frac{\mathbf{x^2_{01}}}{
\mathbf{x^2_{02}}\,\,
{\mathbf{x^2_{12}}} } }
\eea
 $N\Lb Y; \vec{x}_{01}, \vec{b}\Rb$ denotes the dipole scattering 
amplitude. $\vec{x}_{01} = \vec{x}_1 - \vec{x}_0 \equiv \vec{r}$  the
 size of the dipole. The kernel $K\Lb \vec{x}_{01};\vec{x}_{02},
\vec{x}_{12}\Rb$ describes the decay of the dipole with size $x_{01}$
 into two dipoles of  size: $\vec{x}_{02}$ and $\vec{x}_{12} = 
\vec{x}_{01} - \vec{x}_{02}$. The linear part of this equation reduces
 to the BFKL equation\cite{BFKL} with the eigenfunctions
 $\Lb r^2\Rb^\gamma$
which corresponds to the value of the kernel:
\beq \label{KER}
K\Lb \vec{x}_{01};\vec{x}_{02}, \vec{x}_{12}\Rb\,\,\,\longrightarrow\,\,
\, \chi\Lb \gamma\Rb\,\,=\,\,2 \psi(1)\,-\,\psi(\gamma))\,-\,
\psi(1 - \gamma)
\eeq
where $\psi(z) = d \ln \Gamma(z)/d z $ is the Euler $\psi$-function
 (see Ref.\cite{RY} formula {\bf 8.36}).

At first sight, we need to solve this equation to obtain the
 inclusive cross section. However, we need to know the dependence of
 the dipole amplitude on the impact parameter, which cannot be found 
from \eq{BK}.  The failure to reproduce the correct large $b$ behaviour
 of the scattering amplitude, is a long standing problem of 
 non-perturbative QCD  contributions \cite{KOWI} to the BK equation.
  As a consequence we are doomed to use  a phenomenological input in
 addition to \eq{BK}. Hence, we suggest  the following strategy: to 
use a simplified form of \eq{BK}  and to  study in this approach
 the main features of the inclusive production. After such an
 investigation, we will select the model which  satisfies both 
the BK equation, and reproduces the correct behaviour at large $b$.

The BFKL kernel of \eq{KER} includes the summation over all twist
 contributions. In the simplified approach we restrict ourselves 
 to the leading twist term only, which has the form 
\bea \label{SIMKER}
\chi\Lb \gamma\Rb\,\,=\,\, \left\{\begin{array}{l}\,\,\,\frac{1}{\gamma}\,\,\,\,\,\,\,\,\,\,\mbox{for}\,\,\,\tau\,=\,r Q_s\,<\,1\,\,\,\,\,\,\mbox{summing} \Lb \ln\Lb1/(r\,\Lambda_{\rm QCD})\Rb\Rb^n;\\ \\
\,\,\,\frac{1}{1 \,-\,\gamma}\,\,\,\,\,\mbox{for}\,\,\,\tau\,=\,r Q_s\,>\,1\,\,\,\,\,\mbox{summing} \Lb \ln\Lb r Q_s\Rb\Rb^n;  \end{array}
\right.
\eea
instead of the full expression of \eq{KER}.

As  indicated in \eq{SIMKER}  we have two types of logs: $ \Big(\bas
 \ln\Lb r\,\Lambda_{QCD}\Rb\Big)^n$ in the perturbative QCD kinematic
 region where  $r\,Q_s\Lb Y,b\Rb\,\,\equiv\,\,\tau\,\,\ll\,\,1$; and
 $  \Big(\bas \ln\Lb r \,Q_s\Lb Y,b \Rb \Rb\Big)^n$ inside the
 saturation domain ($\tau\,\gg\,1$), where $Q_s\Lb Y, b \Rb$ denotes the
 saturation scale. To sum these logs  it is necessary to modify the 
BFKL kernel
 in different ways in the two kinematic regions, as  shown in 
\eq{SIMKER}.
 
$ \bullet \,\,{\it \tau \,\,=\,\, r\,Q_s \ll\,1}$

For the perturbative QCD region of $\tau \,\ll\,1$, the logs originate 
from
 $ x^2_{02} \sim x^2_{12} \,\ll\, x^2_{01}\equiv r$ resulting in the following
 form of the kernel $ \displaystyle{K\Lb \vec{x}_{01};\vec{x}_{02},
\vec{x}_{12}\Rb}$ \cite{LETU}
 \beq \label{K1}
 \int d^2 x_{02}\, \displaystyle{K\Lb \vec{x}_{01};\vec{x}_{02},
\vec{x}_{12}\Rb}\,\,\,\,\rightarrow\,\pi\, x^2_{01}\,\int^{\frac{1}{\Lambda^2_{QCD}}}_{r^2} \frac{ d x^2_{02}}{x^4_{02}}
\eeq
The non-linear BK equation in this region  can be written as 
\beq \label{BK1}
\frac{\partial^2 n\Lb Y; \vec{x}_{01}, \vec{b}
 \Rb}{\partial Y\,\partial \ln\Lb 1/(x^2_{01}\, \Lambda^2_{QCD})\Rb}\,\,\,=\,\,\frac{\bas}{2}\,\Big( 2 n\Lb Y; \vec{x}_{01}, \vec{b}
 \Rb\,\,- \,n^2 \Lb Y; \vec{x}_{01}, \vec{b}
 \Rb\Big)
\eeq
for $n\Lb Y; \vec{x}_{01}, \vec{b}
 \Rb\,=\,N\Lb Y; \vec{x}_{01}, \vec{b}
 \Rb/x^2_{01}$ .
 
 $ \bullet \,\,{\it \tau \,\,=\,\, r\,Q_s \gg\,1}$ 
 
Inside  the saturation region where $\tau\,\,>\,\,1$ the logs 
   originate from the decay of a large size dipole into one small
 size dipole  and one large size dipole.  However, the size of the
 small dipole is still larger than $1/Q_s$. This observation can be
 translated in the following form of the kernel
\beq \label{K2}
 \int \, \displaystyle{K\Lb \vec{x}_{01};\vec{x}_{02},\vec{x}_{12}\Rb}\,d^2 x_{02} \,\,\rightarrow\,\pi\, \int^{x^2_{01}}_{1/Q^2_s(Y,b)} \frac{ d x^2_{02}}{x_{02}^2}\,\,+\,\,
\pi\, \int^{x^2_{01}}_{1/Q^2_s(Y, b)} \frac{ d |\vec{x}_{01}  -
 \vec{x}_{02}|^2}{|\vec{x}_{01}  - \vec{x}_{02}|^2}
\eeq

Inside the saturation region the BK equation takes the form
\beq \label{BK2}
\frac{\partial^2 \widetilde{N}\Lb Y; \vec{x}_{01}, \vec{b}\Rb}
{ \partial Y\,\partial \ln r^2}\,\,=\,\, \bas \,\left\{ \Lb 1 
\,\,-\,\frac{\partial \widetilde{N}\Lb Y; \vec{x}_{01}, \vec{b}
 \Rb}{\partial  \ln x^2_{01}}\Rb \, \widetilde{N}\Lb Y; \vec{x}_{01},
 \vec{b}\Rb\right\}
\eeq
where 
 $\widetilde{N}\Lb Y; \vec{x}_{01}, \vec{b}\Rb\,\,=\,\,\int^{
 x^2_{01}} d x^2_{02}\,N\Lb Y; \vec{x}_{02}, \vec{b}\Rb/x^2_{01}$ .
 
 The advantage of the simplified kernel of \eq{SIMKER} is 
that,  in the  Double Log Approximation (DLA) for $\tau < 1$,
 it provides a matching with the DGLAP evolution equation\cite{DGLAP}.
 
%%%%%%%%%%%%%%%%%%%%%%%%%%%%%%%%%%%%%%%

\subsection{Solutions}
%%%%%%%%%%%%%%%%%%%%%%%%%%%%%%%%%%%%%%%%

\subsubsection{Perturbative QCD: linear equation}

%%%%%%%%%%%%%%%%%%%%%%%%%%%%%%%%%%%%%%%%%%%%
First we  discuss the solution to \eq{BK1} in 
the  perturbative QCD region, where one can neglect the contributions
 of the shadowing corrections. The equation has the form:
\beq \label{DLA}
\frac{\partial^2 n\Lb Y; \vec{x}_{01}, \vec{b}
 \Rb}{\partial Y\,\partial \ln\Lb 1/(x^2_{01}\, \Lambda^2_{QCD})\Rb}\,\,\,=\,\,\bas\, n\Lb Y; \vec{x}_{01}, \vec{b}
 \Rb
\eeq
One can recognize that \eq{DLA} is the DGLAP equation\cite{DGLAP} in
 the double log approximation (DLA). It has the following solution
 (see for example Ref.\cite{KOLEB})
\beq \label{SOLK1}
N\Lb Y; x_{01}, b \Rb\,\,=\,\,N_0\,\exp\Lb \sqrt{-\,\xi_s\,\xi}\,\,+\,\,\xi\Rb \,\xrightarrow{\tau \to 1;  \zeta \,\to\,0_-} \,\,N_0 e^{\h  z  }\,\exp\Lb - \frac{z^2}{8 \xi_s}\Rb
\eeq
where we use the following notation:
\beq \label{NOTA}
\xi_s\,\,=\,\,4\,\bas\Lb Y - Y_{\rm min}\Rb;~~~~~\xi\,\,=\,\,\ln\Lb x^2_{01}\, Q^2_s\Lb Y=Y_{\rm min}; b\Rb\Rb ;~~~~~~~~ z\,\,=\,\,\xi_s\,+\,\xi;
\eeq
The solution of \eq{SOLK1} provides the boundary condition for the 
solution
 inside the saturation region:
\beq \label{INCK1}
N\Lb Y; z\, =\, 0_- (\xi = -\xi_s), b \Rb\,\,=\,\,N_0\Lb b \Rb;~~~~~~~~~~~~ \frac{\partial \ln  N\Lb Y; z=0_- (\xi = -\xi_s), b \Rb}{\partial \zeta}\,\,=\,\,\h;
\eeq

As  was expected\cite{GS3}, in the vicinity of the saturation scale
 ( $\zeta \ll 8 \xi_s$), the amplitude exhibits  geometric 
scaling behavior, being a function of  only one variable 
$\zeta$  and it has the following form\cite{MUT}:
\beq \label{GSK1}
N\Lb Y; r, b \Rb\,\,=\,\,N_0\, \Lb r^2 Q^2_s\Lb Y, b \Rb \Rb^{ 1 - \gamma_{cr}}
\eeq 

where $\gamma_{cr}$ denotes the critical anomalous dimension which 
in the DLA is equal to $\h$.

%%%%%%%%%%%%%%%%%%%%%%%%%%%%%%%%%%%%%%%%

\subsubsection{  Solution in the region $\tau \,<\,1$}

%%%%%%%%%%%%%%%%%%%%%%%%%%%%%%%%%%%%%%%%%%%%
The solution of \eq{GSK1} assumes that the value of the  constant $N_0$ in
  this equation is small, and we can neglect the non-linear term. However,
 we need to estimate the non-linear corrections in this region, due
 to the value of $N_G$,  which we need to evaluate for the calculation
 of the inclusive production (see \eq{NG}), explicitly accounts for the 
$N^2$
 term. \eq{BK1} can be simplified assuming the geometric scaling behaviour
 of the amplitude in the vicinity of the saturation scale\cite{GS3}. It
 has the form:
\beq \label{BKGS1}
\frac{d ^2 n\Lb z \Rb}{d z^2}\,\,\,=\,\,\frac{1}{8}\,\Big( 2 n\Lb z
 \Rb\,\,- \,n^2 \Lb z
 \Rb\Big)
\eeq
We can obtain the solution to \eq{BKGS1} introducing $d n/d z= p\Lb n\Rb$ 
.
 For this function we obtain the equation:
\beq \label{BKGS11}
\frac{d \, p^2}{d\, n}\,=\,\frac{1}{4}\Big( 2 n\,\,- \,n^2 \Big)
 \eeq
 which has the solution
 \beq \label{BKGS12}
 p^2\,\,=\,\,\frac{1}{4}\Big(  n^2\,\,- \,\frac{n^3}{3} \Big) + {\rm C}_1
\eeq

Since $p \propto n$ for small $n$, as we have seen above, we conclude that
 $C_1 = 0$.
The amplitude $n$ can be found by solving the algebraic  equation:
\beq \label{BKGS13}
2 \int^n_{n_0} \frac{d n' }{n'} \frac{1}{\sqrt{1 - n/3}}\,\,=\,\, z\,\,\,
=\,\,\,2 \Bigg(\ln \frac{ \left(1-\sqrt{1-\frac{n}{3}}\right)}{\left(1 +
 \sqrt{1-\frac{n}{3}}\right)} \,\,-\,\,\ln \frac{ \left(1-\sqrt{1-\frac{n_0}{3}}\right)}{\left(1 + \sqrt{1-\frac{n_0}{3}}\right)}\Bigg)
\eeq
The first correction to the scattering amplitude of  order  $n^2$ has
 the following form:
\beq \label{BKGS14}  
n\Lb z \Rb\,\,=\,\,n_0 \,e^{\h \,z\,\,-\,\, \frac{n_0}{6}\Big( e^{\h \,z} \,-\,1\Big)}
\eeq

%%%%%%%%%%%%%%%%%%%%%%%%%%%%%%%%%%%%%%%%

\subsubsection{  Solution in the region $\tau \,>\,1$}

%%%%%%%%%%%%%%%%%%%%%%%%%%%%%%%%%%%%%%%%%%%%

In the saturation region it has been shown \cite{GS1} that the scattering
 amplitude manifests  geometric scaling behavior, which is  supported
 by the experimental data\cite{GS2}. Therefore, \eq{BK2} can be re-written
 in the form:
\beq \label{BKGS2}
\frac{d^2 \widetilde{N}\Lb z \Rb}{  d z^2 }\,\,=\,\, \frac{1}{4} \, \Lb 1 \,\,-\,\frac{ d  \widetilde{N}\Lb z \Rb}{ d \,z} \Rb\, \widetilde{N}\Lb z\Rb
\eeq
Introducing 
\beq \label{PHI}
\frac{ d  \widetilde{N}\Lb z \Rb}{ d \,z}\,\,=\,\,1\,\,-\,\,e^{ - \phi\Lb \widetilde{N} \Rb}
\eeq
we re-write \eq{BKGS2} in the form:

\beq \label{SOLBKGS21}
\frac{d \,\phi\Lb  \widetilde{N}\Rb}{ d \,\widetilde{N}}\frac{d  \widetilde{N}\Lb z \Rb}{ d \,z}\,=\,\,\frac{1}{8}  \widetilde{N} ~~~~~\longrightarrow~~~~\frac{ d\, \phi}{d \,\widetilde{N}}\Bigg( 1 \,\,-\,\,e^{ - \phi}\Bigg)\,\,=\,\, \frac{1}{4} \, \widetilde{N}
\eeq
Integrating  we obtain
\beq \label{SOLBKGS22}
\phi \,\,+\,\,e^{- \phi} \,\,-\,\,1\,\,=\,\,\frac{1}{8} \widetilde{N}^2 \,\,-\,\,{\rm C}
\eeq
which can be resolved as
\beq \label{SOLBKGS23}
2 \sqrt{2} \sqrt{\phi \,\,+\,\,e^{- \phi} \,\,-\,\,1\,\,+\,\,{\rm C}}\,\,=\,\, \widetilde{N}\,\,=\,\,\int^z_0  d z'  \Bigg(1 \,\,-\,\, e^{ - \phi(z')}\Bigg)
\eeq
Taking the derivatives on  both sides of \eq{SOLBKGS23} we reduce
  this equation to the form:
\beq \label{SOLBKGS24}
\sqrt{2}\frac{ \phi'_z}{ \sqrt{\phi \,\,+\,\,e^{- \phi} \,\,-\,\,1\,
\,+\,\,{\rm C}}}\,\,=\,\,1
\eeq
Therefore, we can find $\phi$ as the solution to the following equation:

\beq \label{SOL}
\sqrt{2}\int^{\phi}_{\phi_0} \,d \phi'\,\,\frac{  1}{ \sqrt{\phi' \,\,+\,\,e^{- \phi'} \,\,-\,\,1\,\,+\,\,{\rm C}}}\,\,=\,\,z
\eeq

The value of $C$ has to be found from matching with the region
 $\tau \,<\,1$. For small $\phi_0$ \,\,$C=0$. Indeed, in this
 case the solution at small $\phi$   has the following form: 
\beq \label{SOLSPHI}
\phi\,\,=\,\,\phi_0 \,e^{\h \,z}
\eeq
which coincides with \eq{GSK1} for the region $\tau\,<\,1$ at
 small $\phi_0 \,=\,N_0\,\ll\,1$.

For the values of $N_0$ , which are not very small,  we need to
  find the value of $C$ from the matching conditions, that can
 be taken from  \eq{BKGS14}:
\bea \label{IC}
&&N|_{z= 0_-} \,=\,N_0 = N|_{z= 0_+} \,=\,1 - e^{-\phi_0};\,\,\frac{d N}{ d z}|_{z=0_-} \,=\,\h\,N_0\Big( 1 - \frac{N_0}{6}\Big)\,\,=\,\,\frac{d N}{ d z}|_{z=0_+}\,\,=\,\,\sqrt{\frac{\phi_0 \,\,+\,\,e^{- \phi_0} \,\,-\,\,1\,\,+\,\,{\rm C}}{2}};\nn\\
&&C \,\,=\,\,\,2\,\Bigg(\h\,N_0\Big( 1 - \frac{N_0}{6}\Big)\Bigg)^2 \,\,+\,\,N_0 \,\,+\,\,\ln\Lb 1 - N_0\Rb\,\,\xrightarrow{N_0 \ll 1}\,\,\,- \frac{N_0^3}{2};
\eea
%%%%%%%%%%%%%%%%%%%%%%%%%%%%%%%%%%%%%%%%

\subsubsection{  Inclusive production}

%%%%%%%%%%%%%%%%%%%%%%%%%%%%%%%%%%%%%%%%%%%%

\begin{figure}[ht]
\begin{tabular}{c c }
 \includegraphics[width=7cm]{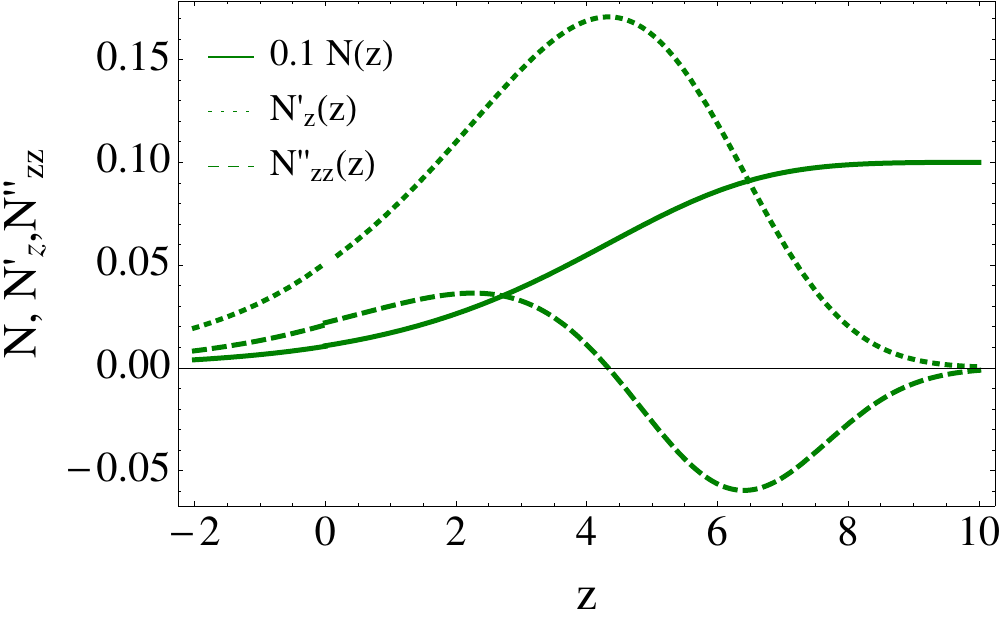}& \includegraphics[width=7.5cm]{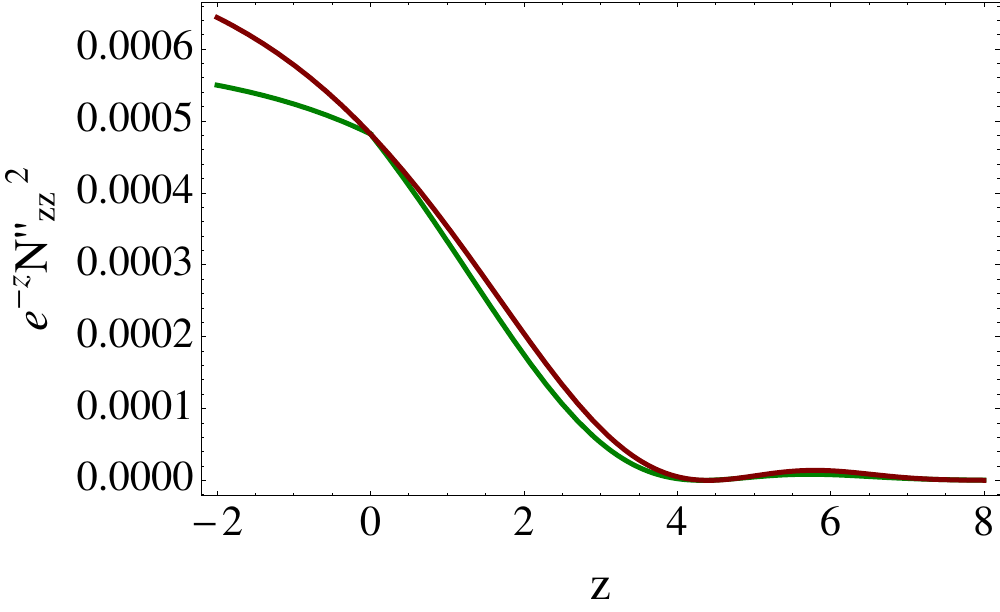}\\
 \fig{ampmod}-a &  \fig{ampmod}-b\\
 \end{tabular}
    \protect\caption{ \fig{ampmod}-a:  numerical solutions of \eq{SOL}
 to the BK equation for small $N_0$ ($N_0 = 0.05$). The impact parameter
 dependence is taken in the form $\Theta\Lb R - b\Rb$ where $\Theta$ is
 the step function. \fig{ampmod}-b: function $I\Lb z\Rb$ in \eq{MF1} (the
 green line) and function
 $e^{-z}\Lb \Big( 1 - e^{- 2 \phi\Lb z\Rb}\Big)''_{z z}\Rb^2$ with $\phi
 = \phi_0 \,e^{\h z}$ ( the red line).
}
\label{ampmod}
   \end{figure}

In  our approach we can estimate the value of
 function ${\cal I}\Lb \tilde{p}\Rb$ in \eq{MF1}. Taking $N_0 = 0.05$, 
and considering this as a small value, we can use \eq{SOL} with $C=0$. In
 \fig{ampmod}-a we present the results of the numerical estimates for 
the amplitude and its derivatives.
   %%%%%%%%%%%%%%%%%%%%%%%%%%%%%%%%%%%%%%%%%%%%%%%%%%%%%%%%%%%%%%%%%%%%%%%
%%%%%%%%%%%%%%%%%%%%%%%%%%%%%%%%%%%%%%%%%%%%%%%% %%%%%%%%%%%%%%%%%%%%
One can see that $N(z) $ tends to 1 at large $z$, but both $N'_z \,=\,d N/d z$
 and $N''_{\!\!z z}\,=\,d^2N/dz^2$ become very small at large  $z$. The integrand
 in 
\eq{MF1} for the inclusive production is equal to $e^{-z}\,N''_{\!\!z z}(z)$
 and it is shown in \fig{ampmod}-b.  From this figure, we note that 
the
 main contribution stems from $z <0$, where our amplitude has the simple
 form of \eq{GSK1}. It turns out that the behaviour for $z >0$ can be
 approximated by the simple formula for the scattering amplitude 
(see \fig{ampmod}-b):
\beq \label{APR1}
N(z)\,\,=\,\,1\,\,-\,\,\exp\Lb - \,\phi_0 e^{\h \,z}\Rb;
\eeq

Hence, one can  see that the situation for inclusive production looks
 quite different  to that for the description of the DIS 
structure function.
  $\nabla^2_T$ in \eq{MF1} generates  an extra factor $e^{-z}$ which
 is large at $ z <0$, and enhances the contribution of the perturbative
 QCD region. In principle,  we expected that  for $p_T \,<\,Q_s$ the
 region of large distances will contribute, and the physical observable
 will be sensitive to the theoretical expectations in this region.
 \fig{ampmod}-b shows that it is not the case for inclusive
 production. To illustrate the influence of the saturation region ($z>0$),
   we calculate the function ${\cal I}\Lb \tilde{p}\Rb$ at $p_T=0$. In 
 ${\cal I}\Lb \tilde{p}_T = 0\Rb$  we expect the largest contribution
  to come from the saturation region. We found, using the numerical 
solution for
 the scattering amplitude, that the contribution of the perturbative
 QCD region at $z <0$ to ${\cal I}\Lb \tilde{p}_T = 0\Rb$ gives 85\%,
 while only 15\% comes from the saturation domain with $z > 0$, for
 the  range $(-8 < z < 8)$.  One can see that the
 integral diverges at $z \to - \infty$, so, we need to generalize
  the behaviour of $\phi$  
 in  the perturbative QCD region (see \eq{SOLK1}) by
 replacing 
\beq \label{REPL}
\phi\,=\,\phi_0 \,e^{\h \,z} \,\,\longrightarrow\,\,\,\phi_0 
\,e^{\hat{\gamma} \,z} 
\eeq
with
\beq \label{NEWG}
\hat{\gamma} \,\,=\,\,\h\,\,-\,\,\frac{z}{8 (4 \bas \ln W))}
\eeq
 \eq{NEWG} is \eq{SOLK1} re-written in the explicit form.

Taking  $\bas = 0.2$ we find that integration over negative
 $z$ gives 85\% of the total contribution.

Hence, we conclude that the following expression provides a
 good approximation  of  the function $I( z )$ in \eq{MF1}:

\bea \label{I}
I\Lb z \Rb\,\,=\,\, e^{-z} \left\{\begin{array}{l}\, \Lb \h N_0 e^{ \h \,z} \,-\, N^2_0 \,e^z\Rb^2,\,\,\,\mbox{for}\,\,\,\tau\,=\,r Q_s\,<\,1\;\\ \\
\,\Lb \Lb1\,\,-\,\,\exp\Lb - 2\,\phi_0 e^{\h \,z}\Rb\Rb''_{\!\!z z}\Rb^2,\,\,\,\, \mbox{for}\,\,\,\tau\,=\,r Q_s\,>\,1;   \end{array}
\right.
\eea

%%%%%%%%%%%%%%%%%%%%%%%%%%%%%%%%%%%%%%%%%%%%%%%%%%%%

The solution $\phi = \phi_0 e^{\h \,z}$ is correct only at small
 values of $N_0$. If $N_0$ is not very small (say is about 1/3) we
 need to take the integral over $\phi'$ in \eq{SOL} keeping $C =
 - N^3_0/2$ in the dominator(see \eq{IC}). The solution for
 $\phi(z)$ at small $z$ has the form
\beq \label{APPR2}
\phi_2(z)\,\,=\,\,N_0\,\Lb 1 - \frac{N_0}{4}\,\Rb e^{\h\,z}\,\,+\,\,\frac{N^2_0}{4} e^{ - \h\,z}
\eeq
One can see that $\phi_2(z) $ is close to $\phi(z)=N_0\,e^{\h\,z}$, even
 for $N_0 = 0.3 - 0.4$.

 We wish to  emphasize that the contribution
 to ${\cal I}\Lb \tilde{p}_T\Rb$ at sufficiently short distances
 does not lead to the suppression of this function at $\tilde{p}_T \,=\, 0$.
 Therefore, the cross section is still  divergent at $p_T \to 0$, which
 results in the large production of soft gluons in the framework of 
CGC/saturation
 approach, and is suppressed by the process of hadronization.

%%%%%%%%%%%%%%%%%%%%%%%%%%%%%%%%%%%%%%%%%%%%%%%%%%%%
\section{Impact-parameter dependent CGC dipole model}

%%%%%%%%%%%%%%%%%%%%%%%%%%%%%%%%%%%%%%%%%%%%%%%%%%%%

 The result of the  assay in the previous
 section  can be formulated as follows: in inclusive production 
the main contribution comes from the vicinity of the saturation scale,
 or in the region of perturbative QCD, while  contributions from
 long distances can be neglected. The behaviour of the amplitude in
 the vicinity of the saturation momentum is predicted theoretically,
 and has the form\cite{MUT}:
\beq \label{VQS}
N\Lb z\Rb\,\,=\,\,N_0\Lb r^2 Q^2_s(x, b \Rb^{\bar \gamma}
\eeq
where $\bar{\gamma} \,=\,1 - \gamma_{cr}$ and $\gamma_{cr} \,=\,0.37$
 in the leading order is the solution to the equation\cite{KOLEB}:
\beq \label{GACR}
\frac{d \chi\Lb \gamma_{cr}\Rb}{d \gamma_{cr}}\,\,=\,\,-\,\frac{\chi\Lb \gamma_{cr}\Rb}{ 1\,-\,\gamma_{cr}}.
\eeq
$\chi\Lb \gamma\Rb$ is given by \eq{KER}.  

The advantage of \eq{VQS} is that we can introduce the correct behaviour
 of the amplitude at large impact parameter by   imposing   the 
phenomenological
   decrease in  saturation momentum for large $b$, by writing it 
in the form:
\beq \label{QSB}
Q_s\,\,=\,\,Q_s\Lb x \Rb S\Lb b \Rb\,\,=\,\,Q_0 \Bigg(\frac{1}{x}\Bigg)^\lambda \,S\Lb b \Rb
\eeq
In the LO BFKL \cite{KOLEB}     $\lambda\,\, 
=\,\,\bas\, \frac{\chi\Lb \gamma_{cr}\Rb}{\bar{\gamma}}$.
Parameters $N_0$ and $Q_0$, as well as function $S\Lb b\Rb$
 should in future
 be taken  from non-perturbative QCD calculations but, at
 the moment, has to be determined from a fit to experimental 
 DIS data.
 We have two models\cite{RESH,CLP}
  \footnote{  The DIS data  has also been described  by the
 saturation model of Ref.\cite{RSKV}, as this model does not produce the
 theoretically correct behaviour deep in the saturation region,  we
 do not consider it in this paper.} 
 on the market that describe the final
 set of the HERA experimental data on deep inelastic structure
 functions\cite{HERADATA} . They have different  forms for $S\Lb b 
\Rb$:
\bea \label{S}
\mbox{Ref\cite{RESH}}: & \to &   S\Lb b \Rb\,\,=\,\,\exp\Lb - \frac{b^2}{B}\Rb\,=\,\exp\Lb - \frac{b^2}{4\,\bar{\gamma}\,B_{\rm CGC}}\Rb;\label{RESHB}\\
\mbox{Ref\cite{CLP}}:& \to &   S\Lb b \Rb\,\,=\,\,\Lb m\,b\,K_1\Lb m\,b\Rb\Rb^{1/\bar{\gamma}};\label{CLPB}
\eea
The ansatz  of \eq{CLPB} is preferable, since it leads to
 $S\Lb b \Rb \xrightarrow{b \gg 1/m} \,\exp\Lb - m b\Rb$,
 which is  in accord with the Froissart theorem\cite{FROI} .
 However, we choose \eq{RESHB} which allow us to do several 
integrations analytically. Using \eq{I} $\nabla^2_T N $ takes 
the following form after integrating over $b$:

\bea \label{I}
\int d^2 b\,\nabla^2_T N \Lb z(r,y,b) \Rb\,\,=\,\, \frac{1}{\tau^2} \left\{\begin{array}{l}\,8\, \pi \, B\, \bar{\gamma }\, N_0\, \tau ^{2 \bar{\gamma }} \Lb  1\,\,-\,\,N_0 \,\tau ^{2 \bar{\gamma }}\Rb; \,\,\,\mbox{for}\,\,\,\tau\,=\,r Q_s(x)\,<\,1\;\\ \\
 8\,\pi\,B\,\bar{\gamma}\,\phi_0 \tau^{2 \bar{\gamma}}\,\,\exp\Lb - 2\,\phi_0  \tau^{2 \bar{\gamma}}\Rb\,\,\,\,\, \mbox{for}\,\,\,\tau\,=\,r Q_s(x)\,>\,1;   \end{array}
\right.
\eea

with $\phi_0\,e^{ - 2 \phi_0}\,=\,N_0\,\Lb 1 - N_0\Rb$.

\begin{figure}[ht]
    \centering
  \leavevmode
      \includegraphics[width=8cm]{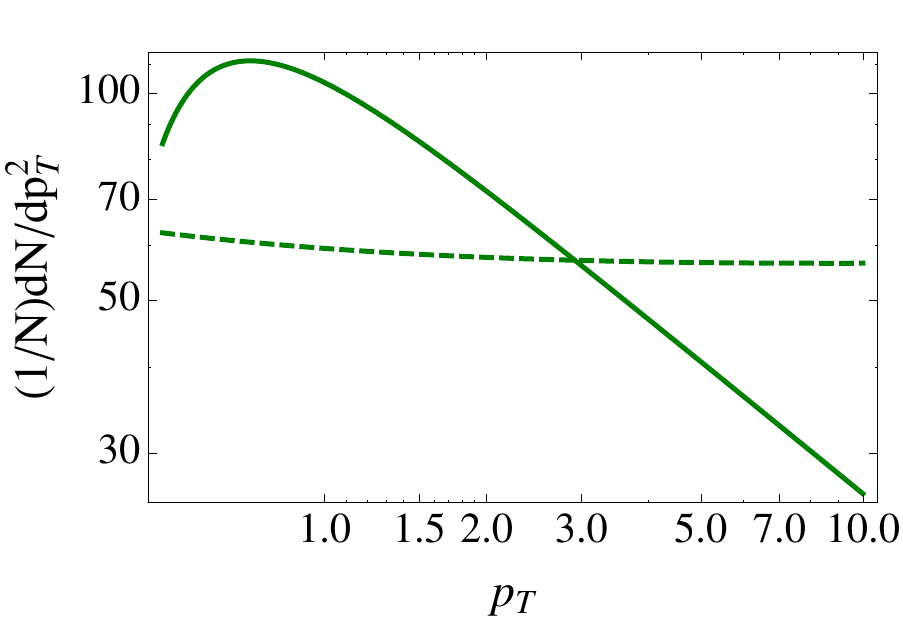}  
    \caption{ The contribution to the gluon inclusive production 
 of \protect\eq{PP}(the solid line)  and of the corrections
multiplied by 100,  due to
 \protect\eq{BKGS13}( the dashed curve).}
\label{cor}
  \end{figure}

Before discussing the details about our model, we would like
 to outline,  which features of \eq{I} stems from the theory,
 and which  from  phenomenological assumptions. The expression 
for $\tau \,\leq\,1$, as we have mentioned (see \eq{VQS})
 follows from the theory. However, the calculation of $N_G$
 (see \eq{NG}) takes into account the term of the order $N^2$.
 As we have demonstrated in section III-B-2 the corrections of
 this order appears in this region in the scattering amplitude
 and has to be included in calculation of $N_G$. We did not take them 
into
 account,  because we view $N = N_0 \tau^{2 \bar \gamma}$
 as a phenomenological expression that describes the DIS
 data\cite{RESH}. However,  we check how large  the contribution
 to the inclusive production from the corrections of \eq{BKGS13}
 and \eq{BKGS14}  is. These corrections are shown  by 
the dashed curve in  \fig{cor}.   One can see that the value of such 
 a correction 
is not more than 2\%.
  %%%%%%%%%%%%%%%%%%%%%%%%%%%%%%%%%%%%%%%%%%%%%%%%%%%%%

 %%%%%%%%%%%%%%%%%%%%%%%%%%%%%%%%%%%%%%%%%%%%%%%%%%%%%%%%%%
The fact that the impact parameter behaviour of the saturation
 momentum determines  the
$b$-dependence of the scattering amplitude comes from theory, while
 the particular form and result of the integration over $b$, stems
 from the model for $S\Lb b\Rb$. 

For $\tau \,\geq\,1$ we have discussed the form of \eq{I} in the 
 previous section, and have given  strong arguments for
 such an expression. The $b$ integration is  performed with the 
phenomenological
 $S\Lb b \Rb$.

In our estimates we use  the values of the parameters from 
Ref.\cite{RESH}(see Table 1). In this paper the HERA data were fitted
in the wide range of $Q^2$ from  $0.75\, GeV^2$ to $650 \, GeV^2$ .
 The expression for $Q_s(x) $ in this model is taken in the
 form\footnote{Note that we introduce the extra factor $\h$
 in the definition of the saturation scale  since we use $\tau =
 r\,Q_s$, while in Ref.\cite{RESH} $\tau$ is defined as $\tau = r Q_s/2$.}
\beq \label{QSX}
Q_s\Lb x \Rb\,\,=\,\,\h \Lb \frac{x_0}{x}\Rb^{\frac{\lambda}{2}}\,GeV
\eeq

 It should be noted, that the value of $x$ from \eq{XF}  even at
  W = 13 TeV, is about $10^{-5}$,  which is in  the region that has been 
measured at HERA.
%%%%%%%%%%%%%%%%%%%%%%%%%%%%%%%%%%%%%%%%%%%%%%%%%%%%%%%%%%%
\begin{table}[h]
\begin{minipage}{13cm}{
\begin{tabular}{|l|l|l|l|l|}
\hline
$\bar{\gamma}$ & $N_0$  & $\lambda$ & $x_0$ & $B_{CGC}$ ($GeV^{-2}$)\\
\hline
 0.6599  $\pm$ 0.0003 & 0.3358$\pm$ 0.0004 & 0.2063 $\pm$ 0.0004 & $0.00105\pm 1.13 10^{-5}$ & 5.5  \\\hline
\end{tabular}
}
\end{minipage}
\begin{minipage}{4cm}
{\caption{Fitted parameters of the model\cite{RESH}, which we use in our
 estimates. }}
\end{minipage}
\label{t1}
\end{table}
%%%%%%%%%%%%%%%%%%%%%%%%%%%%%%%%%%%%%%%%%%%%%%%%%%

\begin{figure}[ht]
\begin{tabular}{c c }
 \includegraphics[width=7cm]{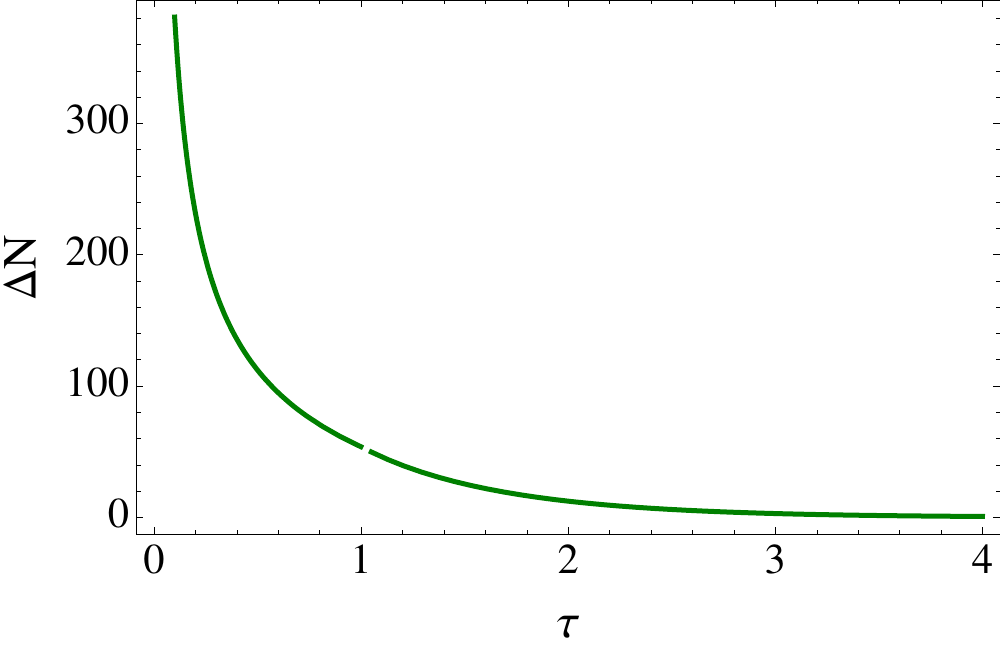}& \includegraphics[width=7.5cm]{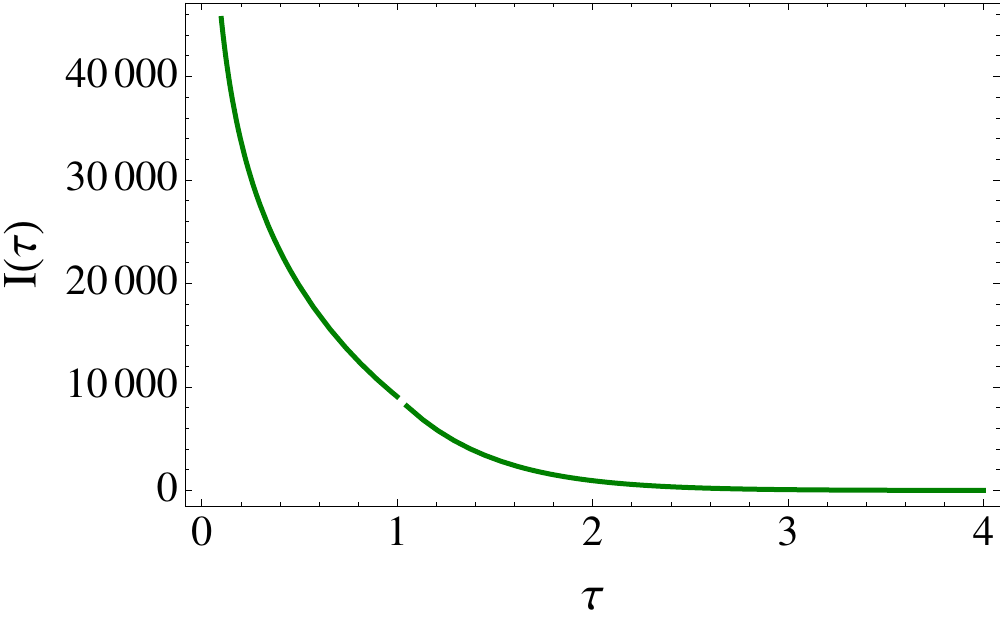}\\
 \fig{modn}-a &  \fig{modn}-b\\
 \end{tabular}
    \protect\caption{ Functions $ \Delta N \equiv 
\int d^2 b\,\nabla^2_T N \Lb z(r,y,b) \Rb $ (\fig{modn}-a)
    and $I\Lb \tau\Rb$ of \eq{MF1}(\fig{modn}-b) versus $\tau$ in
 the model, given by \eq{I}. the values of the parameters are taken
 from Table 1.
}
\label{modn}
   \end{figure}

In \fig{modn} we plot functions 
$\int d^2 b\,\nabla^2_T N \Lb z(r,y,b) \Rb$ and $I\Lb \tau\Rb$ of \eq{MF1}.
 One can see that the both functions provide the main contribution from 
the
 region $\tau \,<\,1$. The largest contribution to the function
 ${\cal I}\Lb\tilde{p}=0\Rb$ in the region of
 integration $\tau\,>\,1$, is  13.5\% . For large $p_T$ we expect 
that 
the contribution of the region $\tau \,>\,1$  to the function ${\cal 
I}\Lb
 \tilde{p}_T\Rb$ is small.  In this region we have an analytical 
expression for this function: 
\beq 
\label{PP}
{\cal I}\Lb \tilde{p}_T\Rb\,\,=\pi ^3\, N_0^2 \,S^2 \,\bar{\gamma}^2\Bigg( \frac{\Gamma\Lb 2 \bar{\gamma} - 1\Rb}{\Gamma\Lb 2 - 2 \bar{\gamma}\Rb}\Lb \tilde{p}_T\Rb^{2 - 4 \bar{\gamma}}\,\,-\,\,N_0\,2^{ 2 \bar{\gamma} +1}\frac{\Gamma\Lb 3 \bar{\gamma} -1\Rb}{\Gamma\Lb 2 - 3 \bar{\gamma}\Rb}\Lb \tilde{p}_T\Rb^{2 - 6 \bar{\gamma}} \,\,+\,\,N^2_0\,2^{2 \bar{\gamma}}\frac{\Gamma\Lb 4 \bar{\gamma} - 1\Rb}{\Gamma\Lb 2 - 4 \bar{\gamma}\Rb}\Lb \tilde{p}_T\Rb^{2 - 8 \bar{\gamma}}\Bigg)
\eeq

   %%%%%%%%%%%%%%%%%%%%%%%%%%%%%%%%%%%%%%%%%%%%%%%%%%%%%%%%%%%%%%%%%%%%%%%
%%%%%%%%%%%%%%%%%%%%%%%%%%%%%%%%%%%%%%%%%%%%%%%% %%%%%%%%%%%%%%%%%%%%

\begin{figure}[ht]
    \centering
  \leavevmode
      \includegraphics[width=8cm]{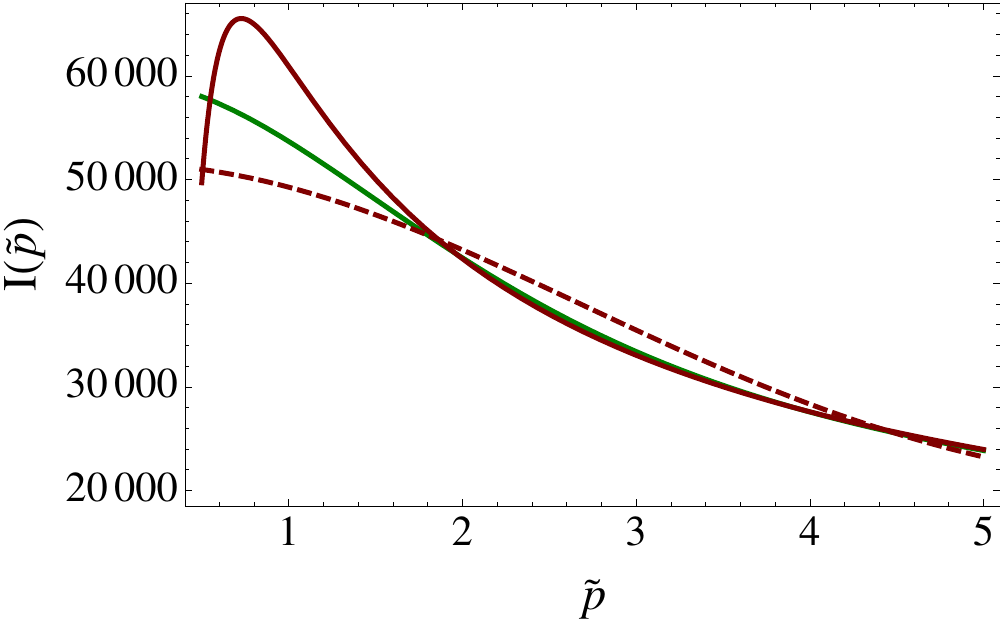}  
    \caption{ The numerical calculations of 
${\cal I}\Lb \tilde{p}_T\Rb$ (see \protect\eq{MF1})
 (the green line) ,  the analytical expressions of \eq{PP} (the
 red solid  line) and of \eq{PR}(the red dashed line)   for this function.}
\label{gg}
  \end{figure}

In \fig{gg} we plot the numerical result for
 ${\cal I}\Lb \tilde{p}_T\Rb$ with functions given by \eq{I}
 and the analytical expression of \eq{PP}. Note that for
  $\tilde{p}_T\, \geq \,2$ (or $ p_T \,\geq \,2 \,Q_s(x)$ )
  both functions coincide.

  %%%%%%%%%%%%%%%%%%%%%%%%%%%%%%%%%%%%%%%%%%%%%%%%%%%%%

 %%%%%%%%%%%%%%%%%%%%%%%%%%%%%%%%%%%%%%%%%%%%%%%%%%%%%%%%%%
  The dashed line in \fig{gg}  illustrates  the result of the 
analytical integration over $\tau$, assuming that the region for $\tau > 
1$ does
 not contribute. The result of this integration has the form:
 
 \bea \label{PR}
&&{\cal I}_{\rm \tau \,\leq\,1}\Lb \tilde{p}_T\Rb\,\,=\,\, \\
&&128 \pi ^3 \gamma ^2 N_0^2  S^2 \left(N_0 \left(\frac{N_0 \, _1F_2\left(4 \bar \gamma -1;1,4 \bar \gamma ;-\frac{\tilde{p}_T^2}{4 }\right)}{8 \gamma -2}+\frac{\, _1F_2\left(3 \gamma -1;1,3 \gamma ;-\frac{\tilde{p}_T^2}{4 }\right)}{1-3 \bar\gamma }\right)+\frac{\, _1F_2\left(2\bar \gamma -1;1,2 \bar\gamma ;-\frac{\tilde{p}_T^2}{4 }\right)}{4 \bar \gamma -2}\right) \nn
\eea
From \fig{gg} we can conclude that  the saturation region with
 $\tau \,\geq \,1$,  only  provides around 10-15\% of the 
contribution
 for $\tilde{p}_T \,\leq\,1$, but the sharp cutoff changes the
 behaviour for large values of $\tilde{p}_T$.

 We know that in the vicinity of the saturation scale
 the scattering amplitude in the momentum representation 
has the following behaviour:
\beq \label{VQSM}
N\Lb p_T\Rb\,\,=\,\,{\rm Const} \Lb \frac{p^2_T}{Q^2_s(x)}\Rb^{\bar{\gamma}}
\eeq
Therefore, from  \eq{PP} we can determine  the value of constant in 
\eq{VQSM} from
 the value of $N_0$.

For large $p_T \,\gg\,Q_s$ we cannot use neither \eq{INCK1} nor
 \eq{VQSM}, since we have to take into account the violation of
 the geometric scaling behaviour in perturbative QCD (see \eq{SOLK1}).
 \eq{SOLK1} indicates that part of this violation can be taken into
 account by replacing in \eq{INCK1}:
\beq \label{REPLACE}
\bar{\gamma}\,\,\longrightarrow\,\,\gamma_{\rm eff} \,\,=\,\,\bar{\gamma} \,\,+\,\,\frac{ \ln(1/\tau)}{\kappa\,\lambda \ln\Lb \frac{1}{x}\Rb }~~~~~~~\mbox{with}~~~~~\kappa\,=\,\frac{\chi''_{\gamma \gamma}\Lb \gamma\Rb}{\chi'_{\gamma }\Lb \gamma\Rb}\Big{|}_{\gamma=\gamma_{cr}}\,\approx \,9.9
\eeq
In the DLA \eq{REPLACE} gives \eq{NEWG}. \eq{REPLACE} is used in
 Ref.\cite{RESH} for fitting the  HERA experimental data.

\begin{figure}[ht]
\begin{tabular}{c c }
 \includegraphics[width=7cm]{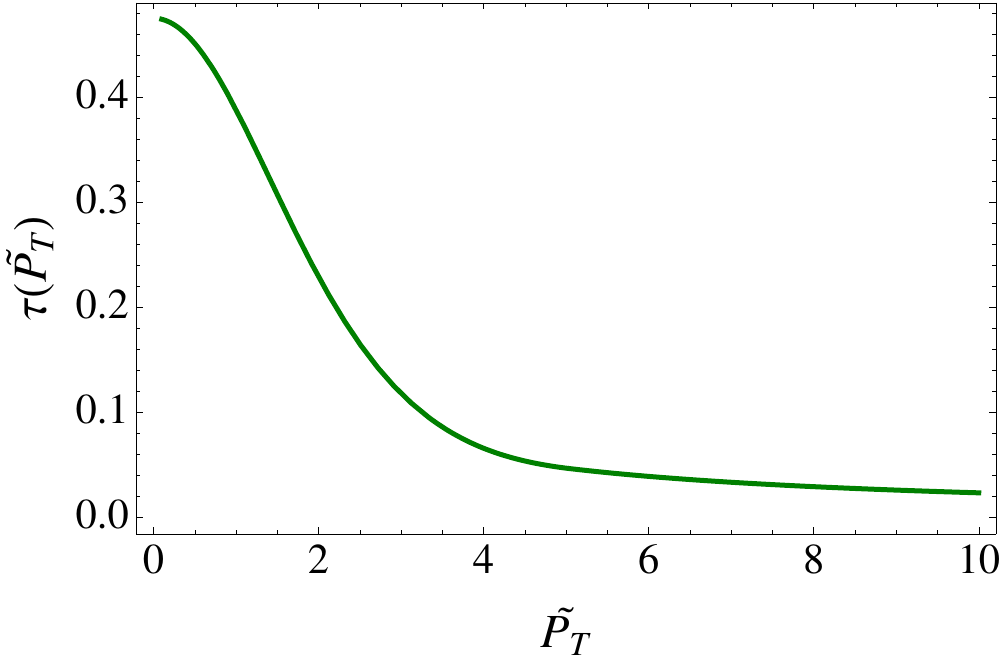}& \includegraphics[width=7.5cm]{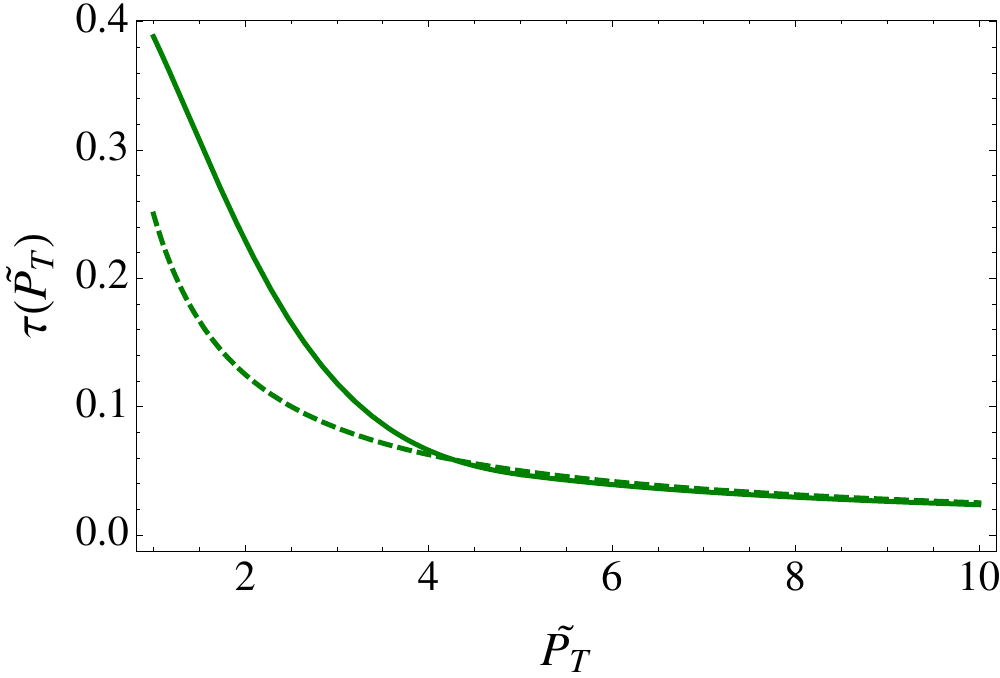}\\
 \fig{tau}-a &  \fig{tau}-b\\
 \end{tabular}
    \protect\caption{ The average $\tau$ (see \eq{AVT}). \fig{tau}-a is the
 result of the numerical calculations.\fig{tau}-b shows that at large
 $\tilde{p}_T$  \,\,$\langle \tau \rangle \, \propto\, 1/(4 \tilde{p}_T)$.}
\label{tau}
   \end{figure}

We make  such a replacement  directly in the momentum
 representation, since $r  \propto 1/p_T$. However, we need to find
 the coefficient in front of $p_T$ and, perhaps, an additional
 constant. We calculate
the average $\tau$  using the expression:
\beq \label{AVT}
\langle \tau \rangle \,=\,\frac{ \int
 \tau J_0\Lb \tilde{p}_T \tau\Rb I\Lb \tau\Rb d \tau}{ \int  J_0\Lb \tilde{p}_T \tau\Rb I\Lb \tau\Rb d \tau}
\eeq
The results of these estimates are shown in \fig{tau}.
 One can see at at 
$\tilde{p}_T \to 0$  
$\langle \tau \rangle\,=\,0.478 \approx \h$ while at
 large $\tilde{p}_T$ it is proportional to $1/(4 \tilde{p}_T)$.

   %%%%%%%%%%%%%%%%%%%%%%%%%%%%%%%%%%%%%%%%%%%%%%%%%%%%%%%%%%%%%%%%%%%%%%%
%%%%%%%%%%%%%%%%%%%%%%%%%%%%%%%%%%%%%%%%%%%%%%%% %%%%%%%%%%%%%%%%%%%%	
  For $\tilde{p}_T=0$, or more generally for $p_T \ll Q_s$,  the
 typical distances turns out to be $r =1/(2 Q_s(x))$, and for large
 $\tilde{p}_T$ they are of the order of $1/(4 p_T)$.
Hence, we suggest to use in \eq{REPLACE}   the calculated $\langle
 \tau \rangle\Lb \tilde{p}_T\Rb$ for $\tilde{p} \leq 4$ and $1/(4
 \tilde{p}_T$) for $\tau \geq 4$.

\begin{figure}[ht]
    \centering
  \leavevmode
      \includegraphics[width=8cm]{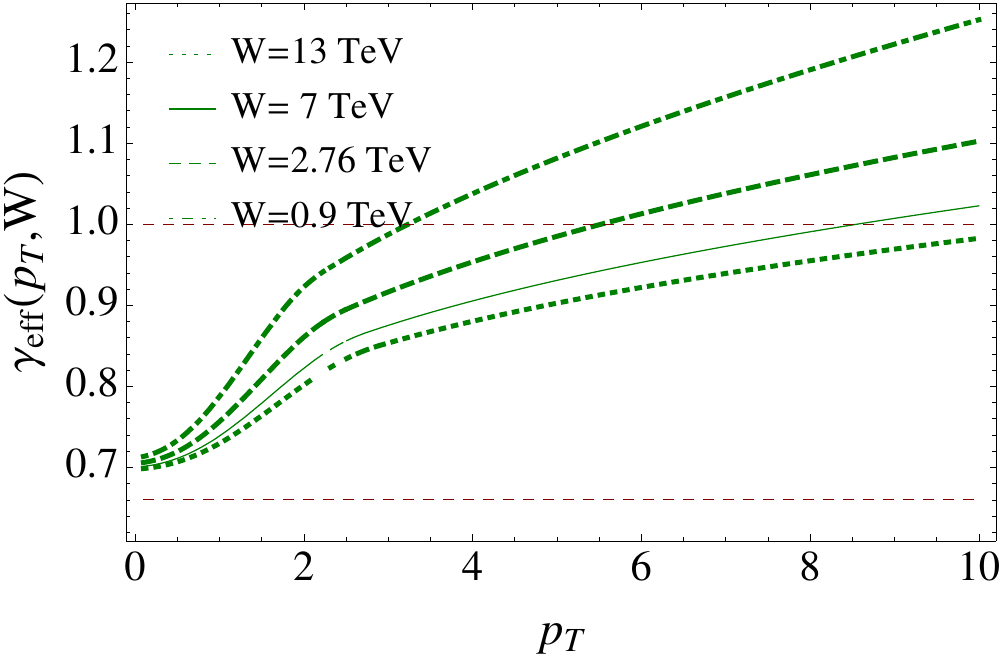}  
    \caption{ The $p_T$  and $W$ dependence  of $\gamma_{\rm eff}$. All
 parameters are taken from Table 1. The dashed red lines show
 $\gamma_{\rm eff} = \bar{\gamma}$ and $\gamma_{\rm eff} =1$.}
\label{gaeff}
  \end{figure}

 In \fig{gaeff}
we show the behaviour of the $\gamma_{\rm eff}$ at different
 energies. One can see that this dependence is essential for
 describing the experimental data. Indeed, the value of $n$
 in the hard term in \eq{SUM} is $n=3.1$. As we  have  seen above
 (see \eq{PP} for example) at large $p_T$ the inclusive cross
 section is proportional to $1/p_T^{4 \gamma_{\rm eff}}$. For
 $\gamma_{\rm eff} = \bar{\gamma} $ it is impossible to obtain
 a decrease of about $1/p^6_T$, as  indicated by the data. 
 The $p_T$ and $W$ dependence of $\gamma_{\rm eff} $ of \eq{REPLACE}
 is shown in \fig{gaeff}.  We  see that even this behaviour of
 $\gamma_{\rm eff}$ leads only to $1/p^{4 -5}_T$ at $p_T \approx $ 7 GeV. 
Hence, we expect that the CGC/saturation approach in the form of
 the model, will not be able to describe the hadron spectra at
 large $p_T$. However, at  large $p_T$ we are outside of the
 vicinity of the saturation scale, and have to perform the
 perturbative QCD calculation which, as we have mentioned,
 describes the experimental data \cite{NLOFIT}.
It should be emphasized, that the discussed problems of the
 CGC approach, has no  influence on the behaviour of the
 transverse momentum distributions at $p_T \leq $ 7 GeV,
 and on the value of the contribution of the thermal
 radiation, which is negligibly small at $p_T \sim 7 $ GeV.

  %%%%%%%%%%%%%%%%%%%%%%%%%%%%%%%%%%%%%%%%%%%%%%%%%%%%%

 %%%%%%%%%%%%%%%%%%%%%%%%%%%%%%%%%%%%%%%%%%%%%%%%%%%%%%%%%%

%%%%%%%%%%%%%%%%%%%%%%%%%%%%%%%%%%%%%%%%%%%%%%%%%%%%
\section{Comparison with the experimental data}

%%%%%%%%%%%%%%%%%%%%%%%%%%%%%%%%%%%%%%%%%%%%%%%%%%%%
As we have discussed in the introduction, our goal is to answer the
 question. Do we need the thermal radiation term to describe the
 experimental data in the framework of the CGC/saturation approach?
  Our answer is yes.

We calculate the cross section for gluon production using \eq{MF1}.
 As we have discussed in function ${\cal I}\Lb \frac{p_T}{Q_s}\Rb$
 the main contribution comes from $p_T \sim Q_s $ (see
\fig{gg}). However, the factor $1/p^2_T$ in front  in \eq{MF1} stems
 from the gluon propagator \cite{BFKL}, and it is affected both by the
 hadronization, and by  interactions with co-movers in the parton
 cascade.  In our approach to the confinement problem, we first 
  need to take into account,   the effect of the mass of produced
 gluon jet due to hadronization, which changes the gluon
 propagator\cite{KHLEKLN}:
\beq \label{GLPROP}
G\Lb p_T\Rb\,\, =\,\, \frac{1}{p^2_T}\,\,\longrightarrow\,\,\frac{1}{p^2_T \,\,+\,\,2 \Lb Q_s \Theta\Lb Q_s - p_T\Rb\,+\,p_T \Theta\Lb p_T - Q_s\Rb\Rb\,m_{\rm eff}}
\eeq
Therefore, we calculate  gluon production using \eq{MF1} in which
 we use \eq{GLPROP} to replace the factor $1/p^2_T$.

\begin{figure}[ht]
    \centering
  \leavevmode
      \includegraphics[width=8cm]{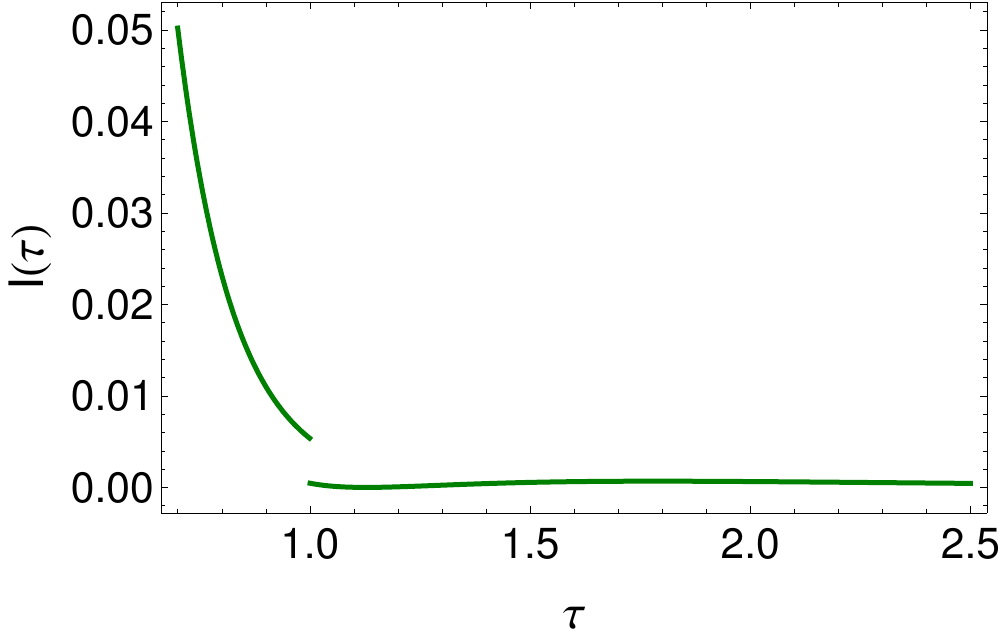}  
    \caption{  The behaviour  function $I\Lb \tau\Rb$ in the model of
 Ref.\cite{RESH}.}
\label{rs}
  \end{figure}

For calculating $I\Lb \tau,W\Rb$ in \eq{MF1} we use
 \eq{I} for $p_T \,\leq\,2\,Q_s(x)$ and \eq{PP} for
 larger values of $p_T$. It should be noted that the original
 model of Ref.\cite{RESH} has a different assumption regarding
 the behaviour of the amplitude in the saturation region.
 However, $\nabla^2 N$ in this function is not continuous
 at $\tau=1$ and  leads to a very small contribution for
 $\tau \geq 1$ (see \fig{rs}).  Hence, in the original
 model we neglect the contribution from the saturation region.
 As we have seen in \fig{gg} such a procedure results in 10-15\%
 accuracy of the calculations  in the entire region of $p_T$.
 It should be mentioned that our assumptions about the behaviour
 of $\nabla^2 N$ in the saturation region, are based on the
 solution to the BK equation in the vicinity of the saturation
 scale. While the model of Ref.\cite{RESH} reproduces only the
the theoretical behaviour of Ref.\cite{LETU} at large $\tau$ or, in 
other words, in the region which does not contribute to $\nabla^2 N$.

  %%%%%%%%%%%%%%%%%%%%%%%%%%%%%%%%%%%%%%%%%%%%%%%%%%%%%

 %%%%%%%%%%%%%%%%%%%%%%%%%%%%%%%%%%%%%%%%%%%%%%%%%%%%%%%%%%

\begin{figure}[ht]
    \centering
  \leavevmode
      \includegraphics[width=8cm]{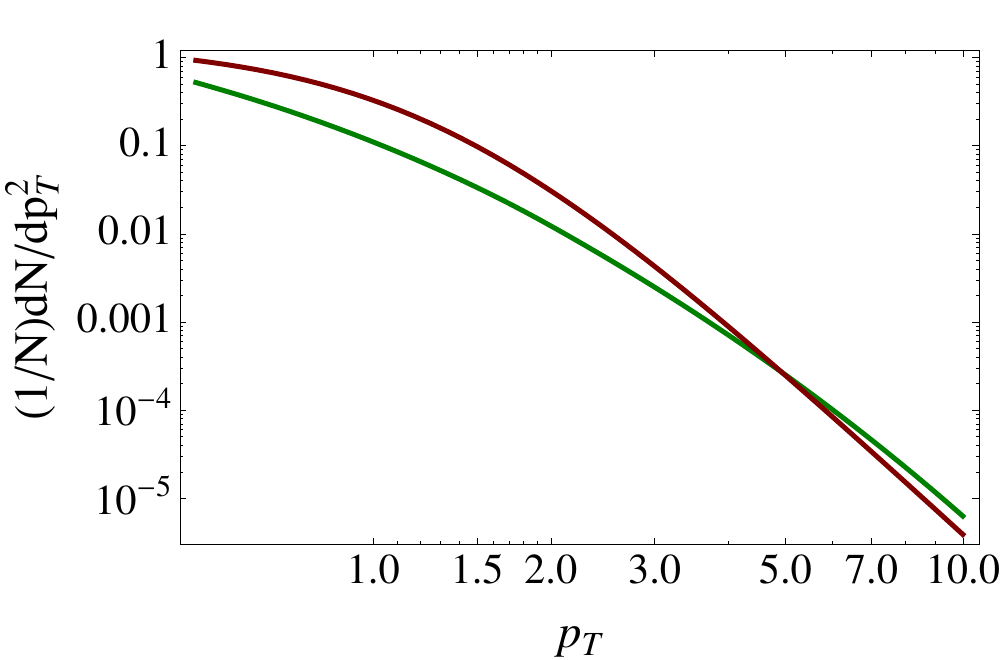}  
    \caption{ The inclusive cross section, calculated in 
CGC/saturation approach compared to the hard term
 in \eq{SUM}. $m_{\rm eff} $= 0.5 GeV.}
\label{htcgc}
  \end{figure}

We  note that the inclusive production calculated from the
 CGC/saturation approach, has  a different form than the hard
 term of \eq{SUM} (see \fig{htcgc}).

  %%%%%%%%%%%%%%%%%%%%%%%%%%%%%%%%%%%%%%%%%%%%%%%%%%%%%

 %%%%%%%%%%%%%%%%%%%%%%%%%%%%%%%%%%%%%%%%%%%%%%%%%%%%%%%%%%
 We have partly  explained that $\gamma_{\rm eff}$ from \fig{gaeff}
 is not able to describe the shape of $p_T$ distribution at high
 $p_T$. On the other hand, we predict the value of the cross section,
 while in \eq{SUM} this value was a fitted parameter.
We will discuss below  the dependence
 of the  cross section on the value of $m_{\rm eff}$.  

\begin{figure}[ht]
\begin{tabular}{c c c}
 \includegraphics[width=6cm]{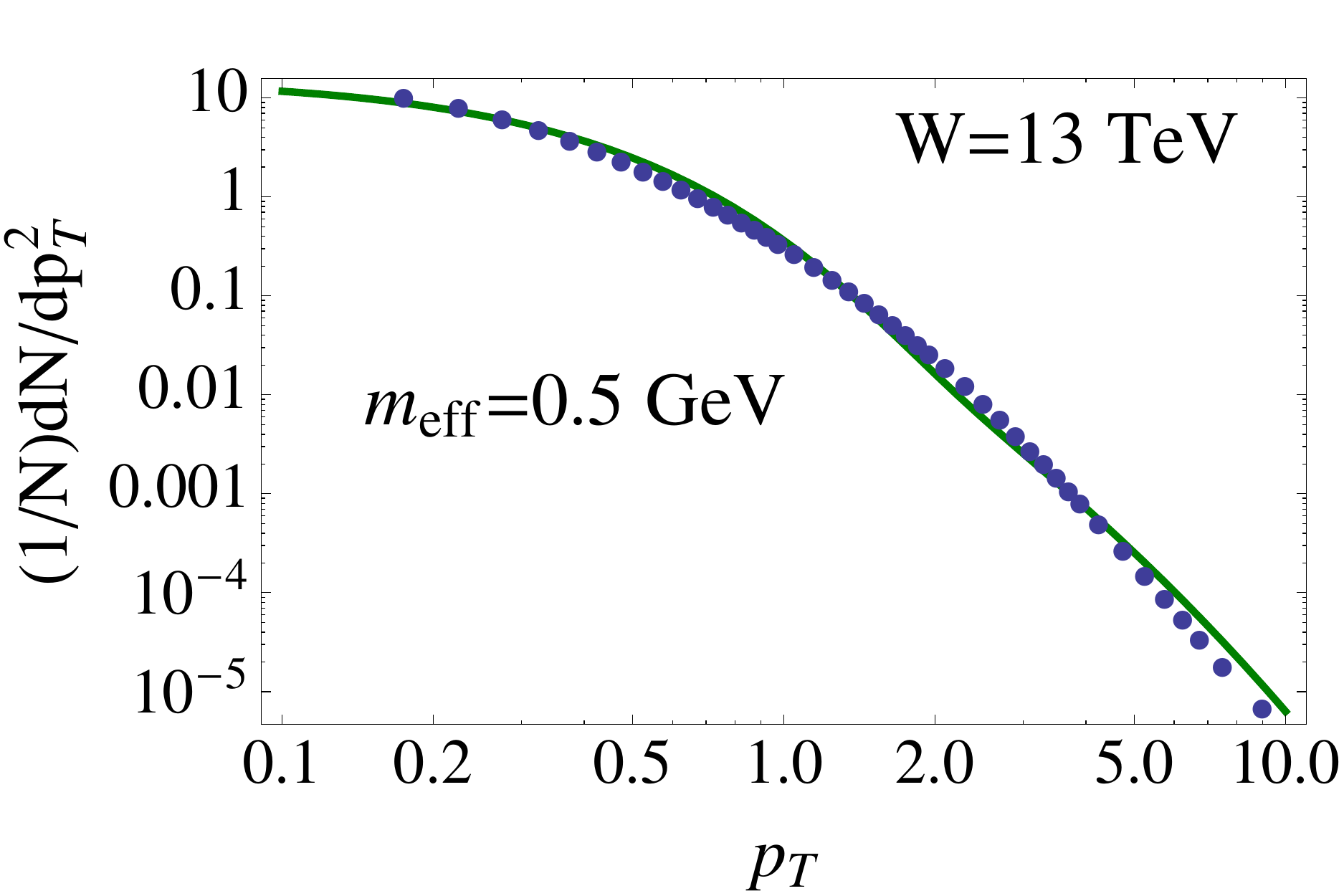}&~~~~& \includegraphics[width=6cm]{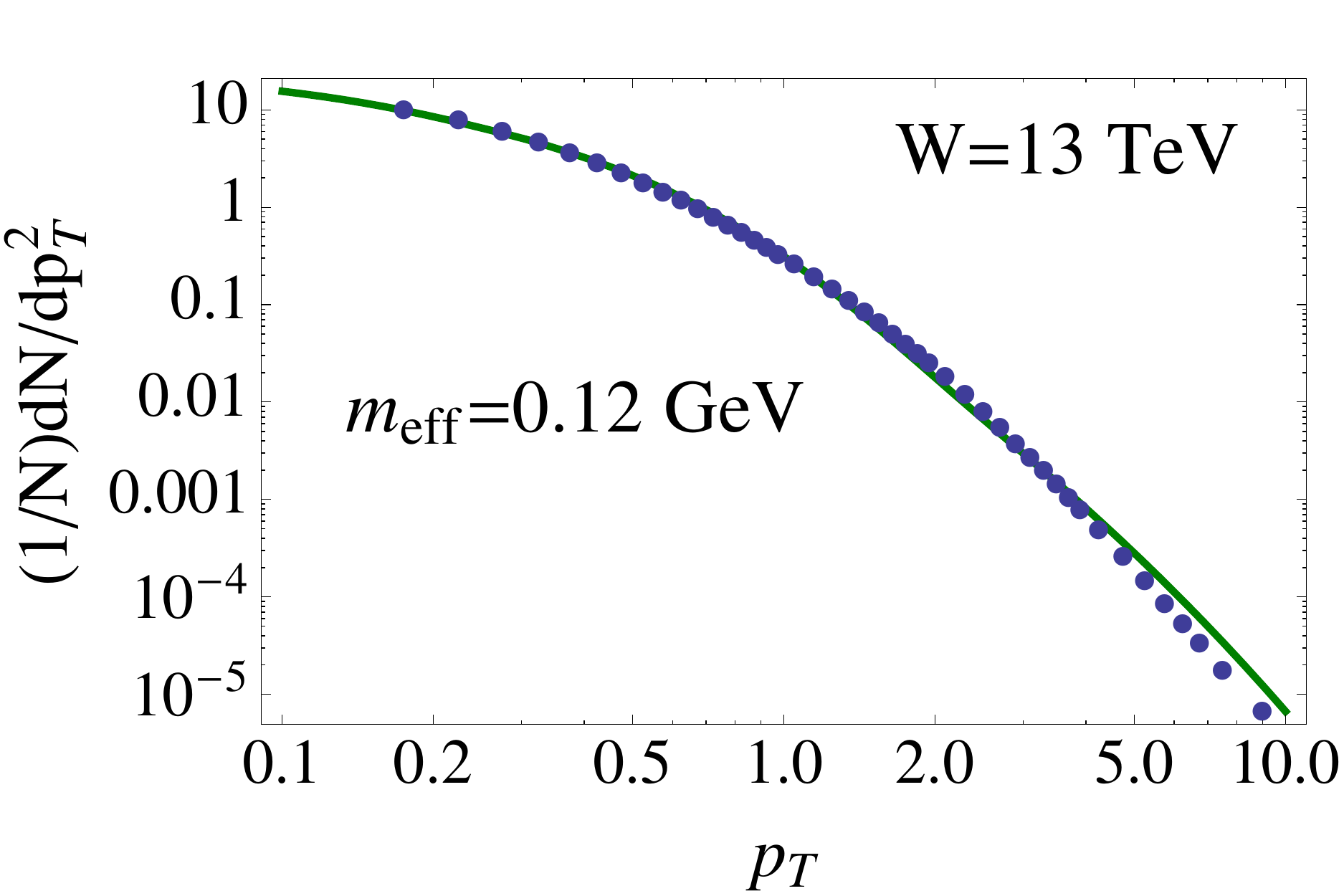}\\
 \fig{fit}-a &  & \fig{fit}-b\\
  \includegraphics[width=6cm]{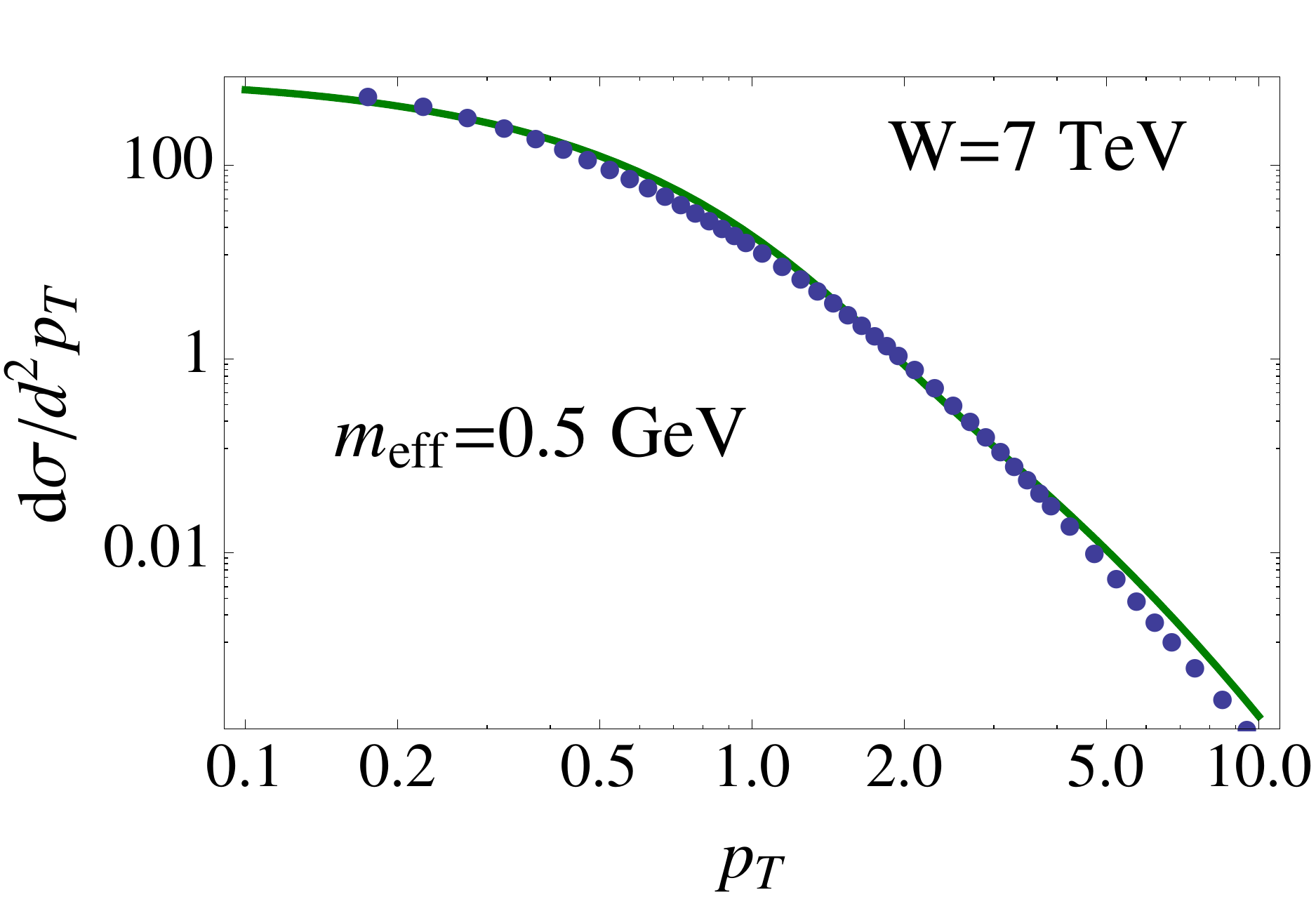}&~~~~& \includegraphics[width=6cm]{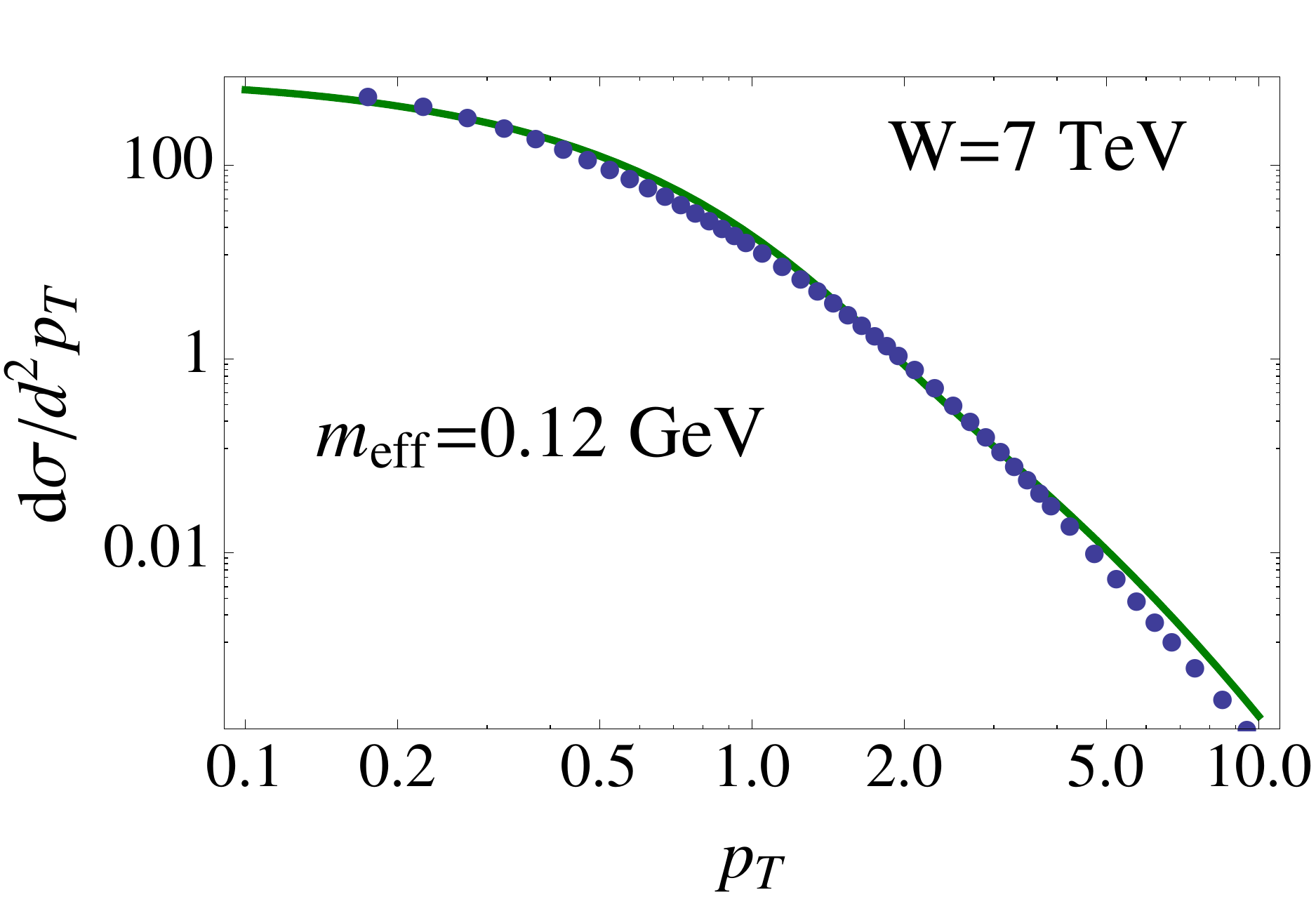}\\
 \fig{fit}-c &  & \fig{fit}-d\\ 
 \includegraphics[width=6cm]{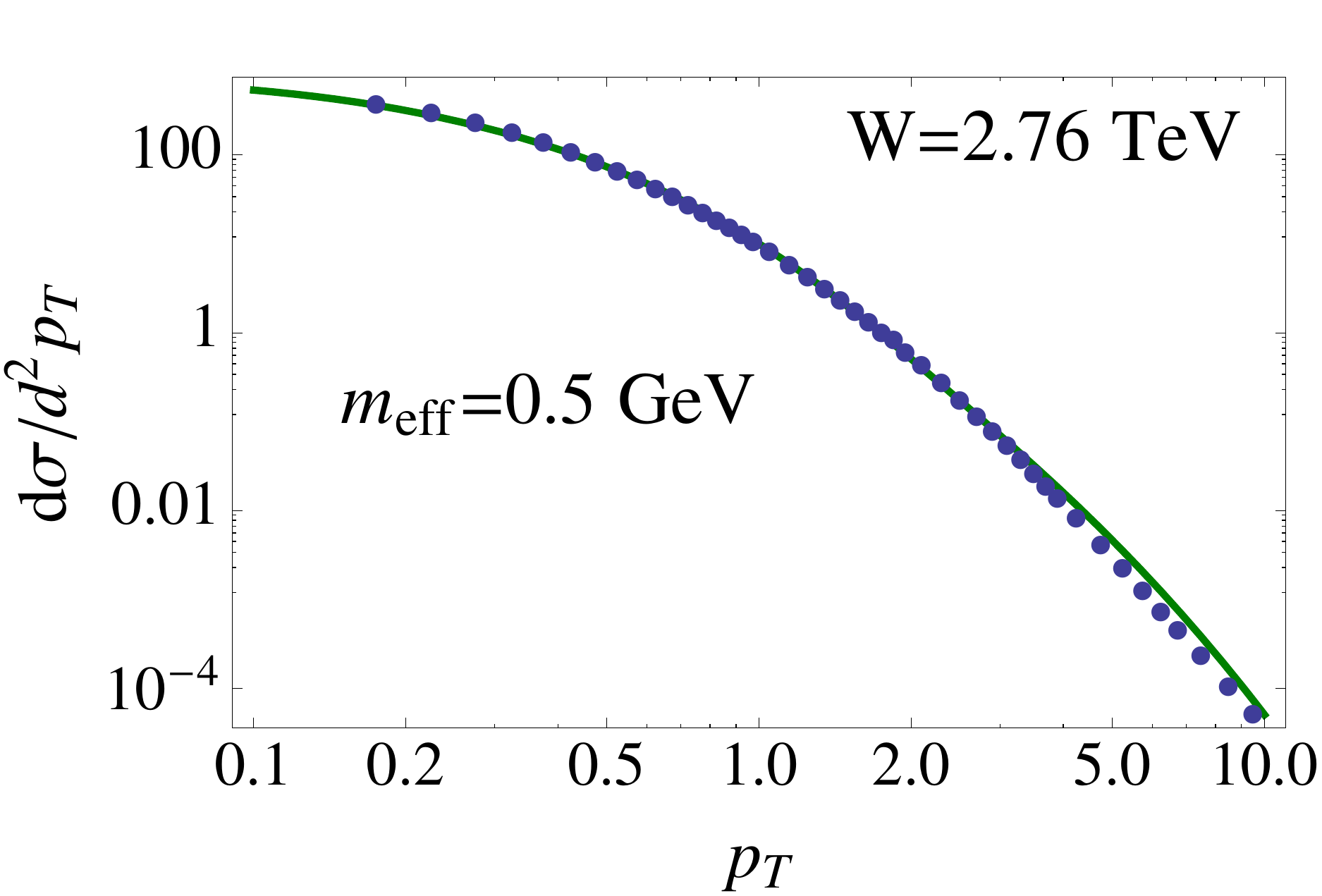}& ~~~~~&\includegraphics[width=6cm]{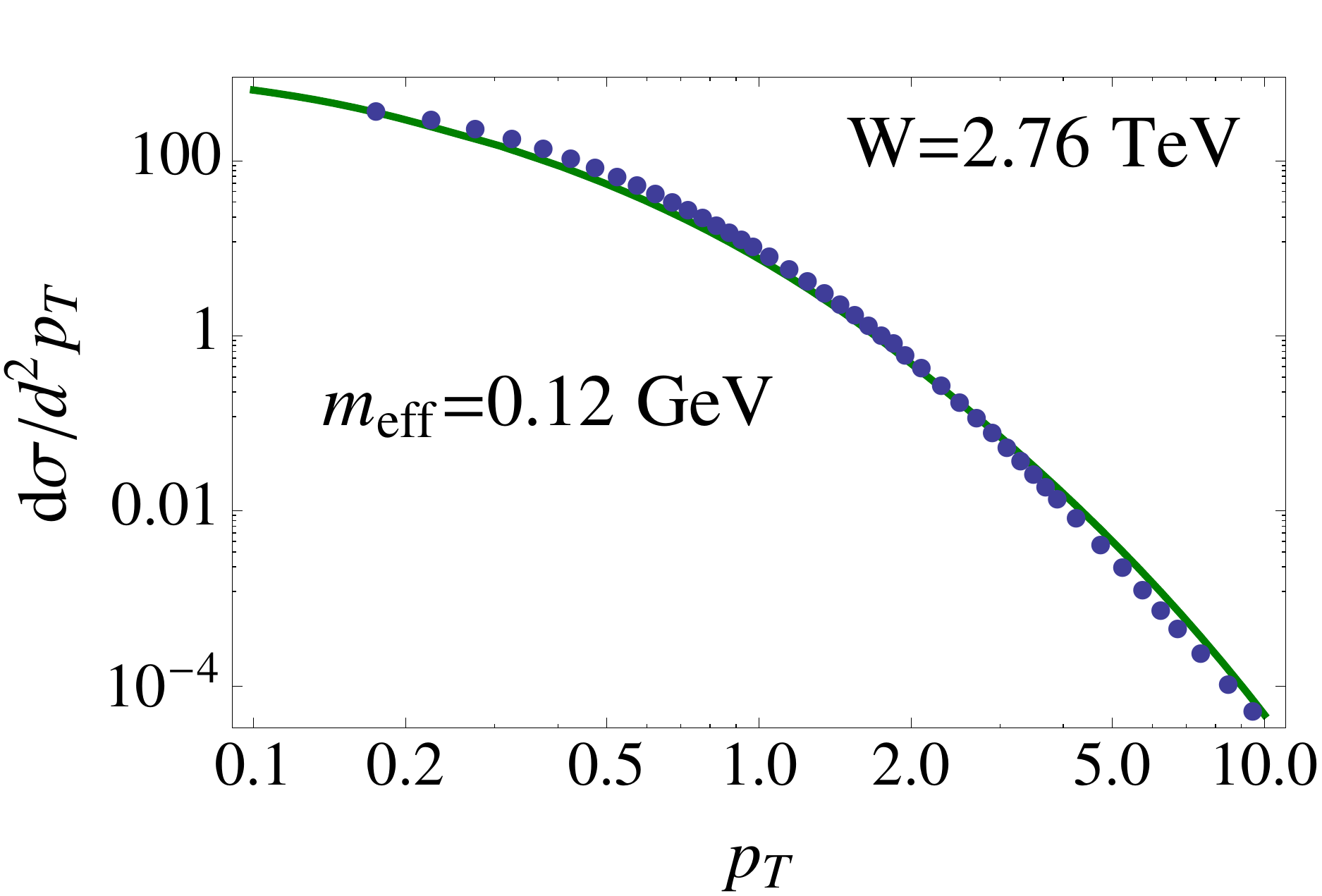}\\
 \fig{fit}-e &  &\fig{fit}-f\\
  \includegraphics[width=6cm]{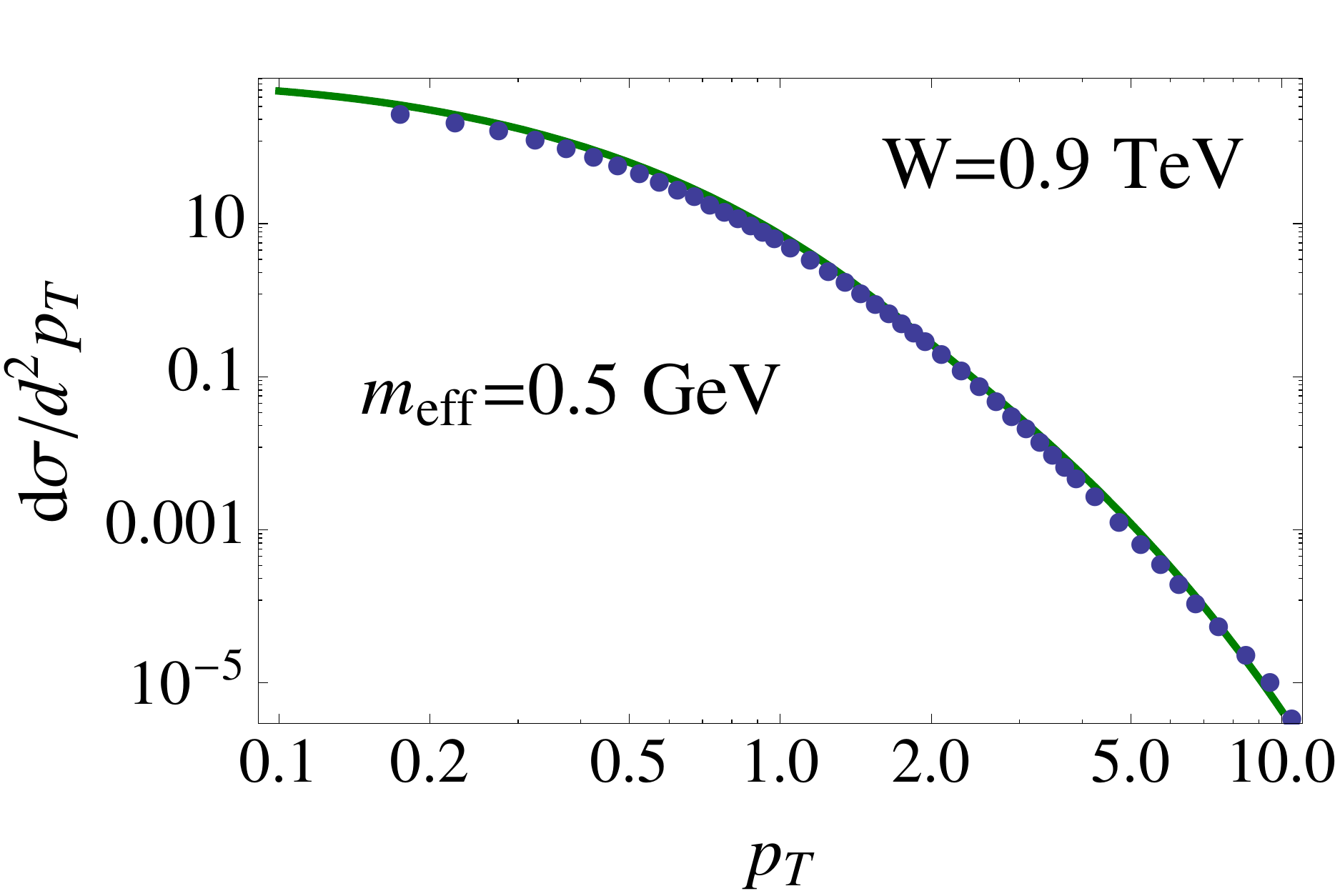}&~~~~~~~& \includegraphics[width=6cm]{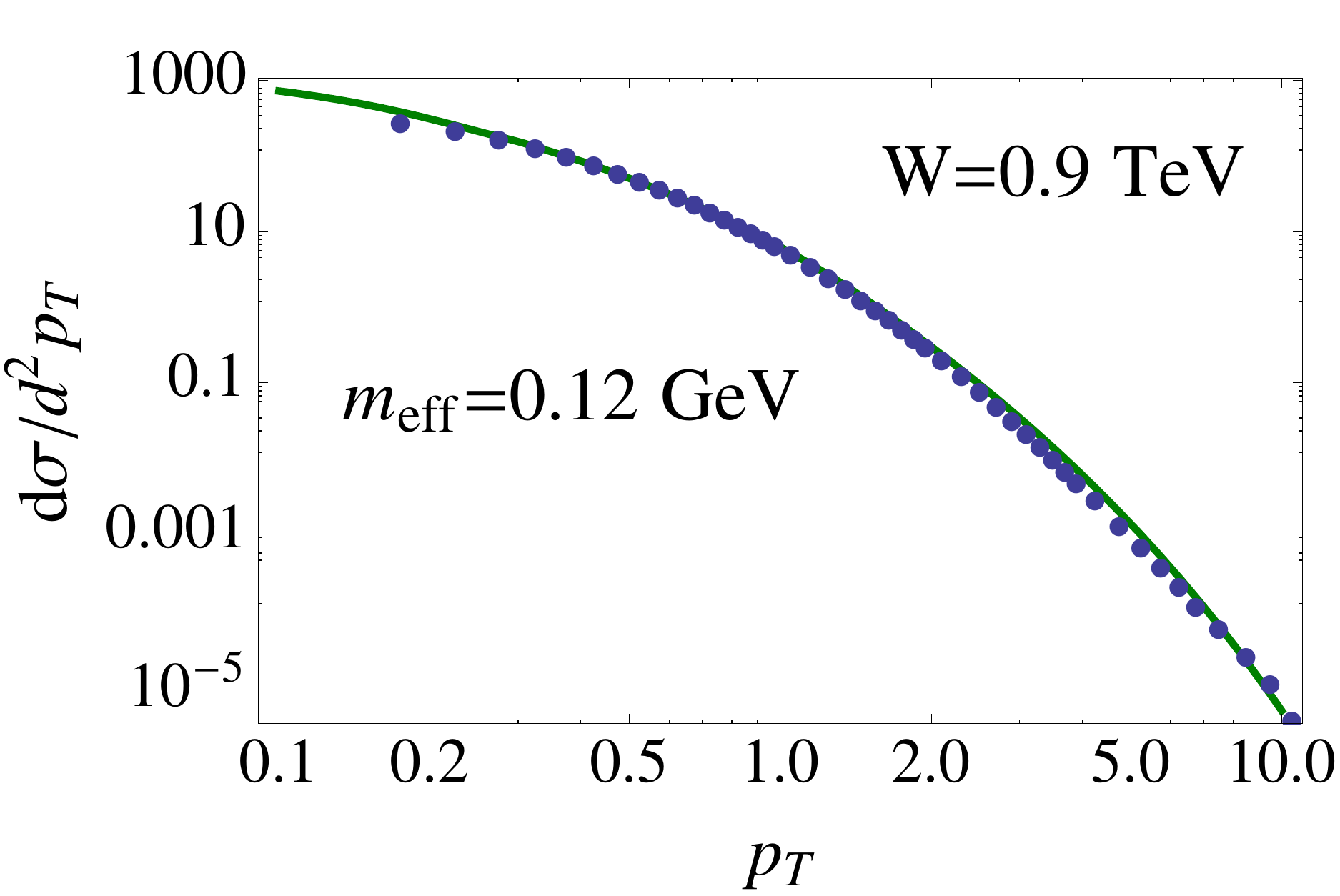}\\
 \fig{fit}-g &  \fig{fit}-h\\ 
 \end{tabular}
    \protect\caption{ The descriptions of the experimental data
 of ALICE collaboration\cite{ALICE13,ALICE2}. For the description of
 the data at W=13 TeV we use the model of Ref.\cite{GLP},  for
 $\sigma_{in}\,\,=\,\,\sigma_{tot} - \sigma_{el} - \sigma_{diff}$. 
 }
\label{fit}
   \end{figure}

 \fig{fit} illustrates how well we describe the data for the
 transverse momentum distribution at the LHC. 

   %%%%%%%%%%%%%%%%%%%%%%%%%%%%%%%%%%%%%%%%%%%%%%%%%%%%%%%%%%%%%%%%%%%%%%%
%%%%%%%%%%%%%%%%%%%%%%%%%%%%%%%%%%%%%%%%%%%%%%%% %%%%%%%%%%%%%%%%%%%%

$T_{\rm th} \,\propto \,Q_S$ was taken from \eq{THTEM} , however, it turns
 out that the experimental data can be described with $c = 2.3$ which
 is almost  twice  larger than estimated in Ref.\cite{KLT}.

\begin{figure}[ht]
\begin{tabular}{c c }
 \includegraphics[width=7cm]{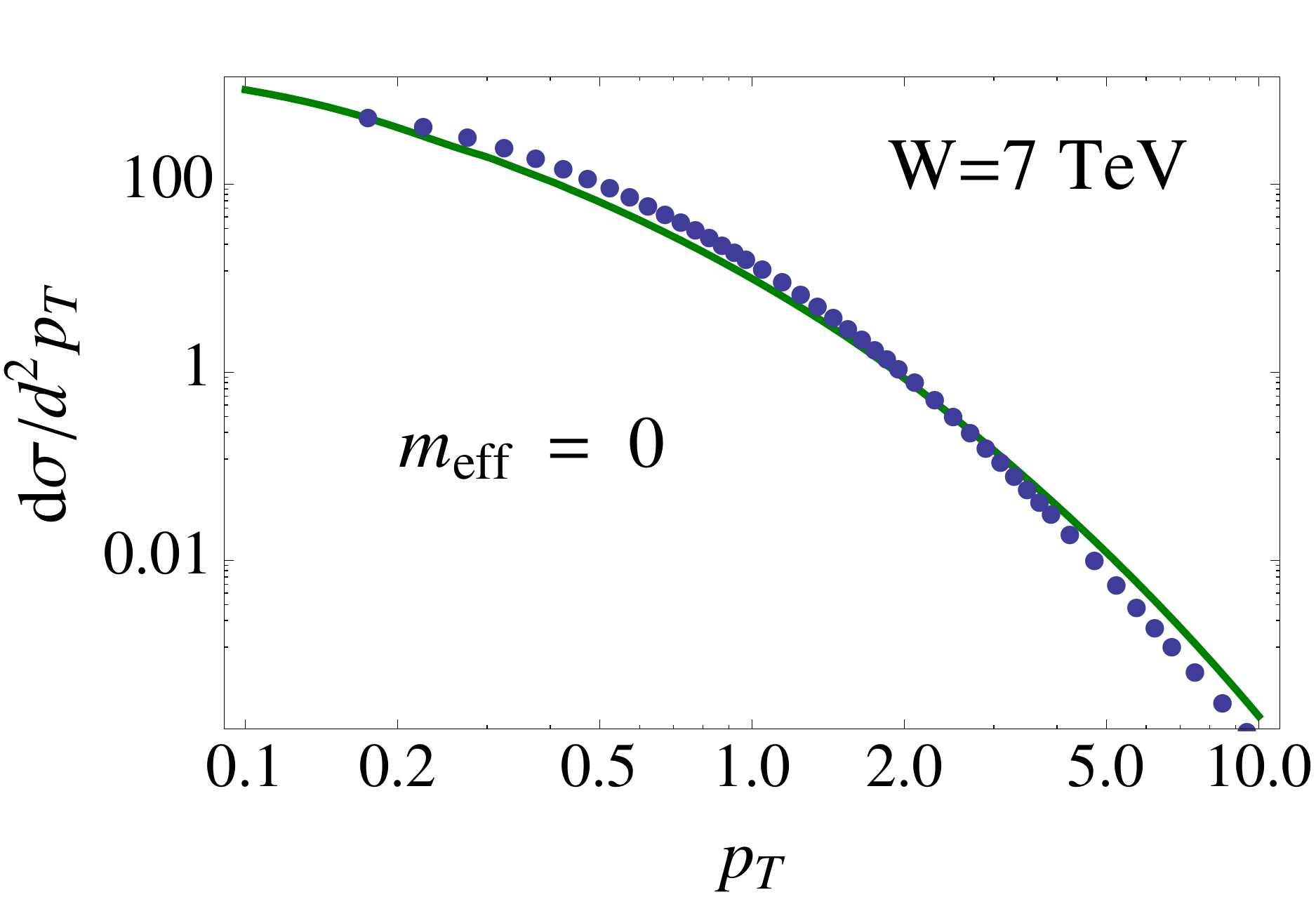}& \includegraphics[width=7cm]{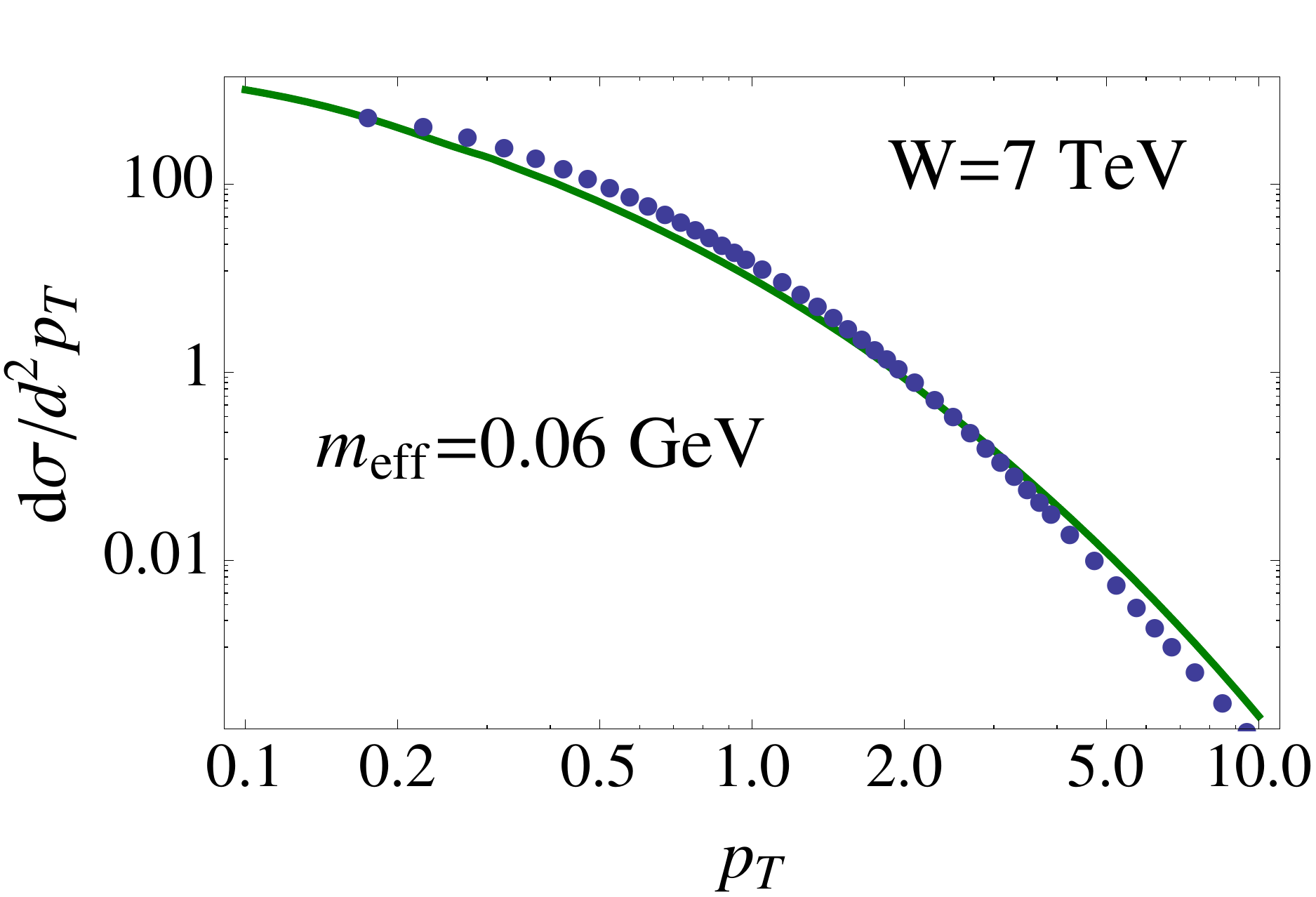}\\
 \fig{meff}-a &  \fig{meff}-b\\
 \end{tabular}
    \protect\caption{ $d^2 \sigma/d^2 p_T$ for W=7 TeV for different
 values of $m_{\rm eff} =0$ (\fig{meff}-a) and $m_{\rm eff}=0.06\,GeV$ without the thermal radiation term.}
\label{meff}
   \end{figure}

 The rate of  thermal radiation is shown in Table 2, in which
 $R = \int  d^2 p_T \,d^2 \sigma^{\rm charged}_{\mbox{therm. rad.}}
 /d^2 p_T/\int  d^2 p_T \,d^2 \sigma^{\rm charged }_{\mbox{sum.}}
 /d^2 p_T$. Note  that the contribution of the thermal
 radiation increases with the growth of energy. The value of
 the CGC term depends on the value of the $m_{\rm eff}$. We
 believe that  most of the pions are produced from  $\rho$
 resonances and we consider $m_{\rm eff} = $ 0.5\, GeV  
 as the most reliable estimate. In \fig{meff} we present
 the calculation at W = 7 TeV  with $m_{\rm eff} = 0$ and
 $m_{\rm eff}=0.06\,GeV$ without the thermal emission term.

%%%%%%%%%%%%%%%%%%%%%%%%%%%%%%%%%%%%%%%%%%%%%%%%%%%%%%%%%%%
\begin{table}[h]
\begin{minipage}{8cm}{
\begin{tabular}{|l|l|l|l|}
\hline
W (TeV)& $m_{\rm eff}$ = 0.5 \,GeV) & $m_{\rm eff}$ = 0.12 \,GeV \\
\hline
13 & 70\% & 46\%    \\\hline
7 & 70\% & 43\%    \\\hline
2.76 & 59\% & 13\%    \\\hline
0.9  & 53\% & 7\%    \\\hline\end{tabular}
}
\end{minipage}
\begin{minipage}{8cm}
{\caption{$R\,\,=\,\,d^2\sigma/d^2p_T (\mbox{thermal radiation})\Big{/}
 d^2\sigma/d^2p_T (\mbox{sum})$ versus the values of energies and the
 value of $m_{\rm eff}$. }}
\end{minipage}
\label{t2}
\end{table}
%%%%%%%%%%%%%%%%%%%%%%%%%%%%%%%%%%%%%%%%%%%%%%%%%%

We see that at small values of the effective mass, we can describe the
 experimental data without the thermal radiation term. It should be
 stressed that we do not need the so called K -factor, to include the
 next-to-leading order corrections. Even for the multiplicity distribution
 at W = 13 TeV we are able to describe the data using
 $\sigma_{in} \,\,=\,\,\sigma_{tot} - \sigma_{el} - \sigma_{diff}$
 from Ref.\cite{GLP}.  We recall that the simple formula for $m_{\rm 
 eff}= \sqrt{\mu^2 + k^2_T +k^2_L} - k_L$ leads to
 $m_{\rm  eff} = 0.5\,GeV$  if $\mu$ is equal to the
 mass of $\rho$-resonance since the value of $k_T = k_L
  = 0.45\,GeV$ (see Ref.\cite{ATLASJET} for the measurement
 and Ref.\cite{KLRX}) and reference therein for theoretical
 discussions). For the minimal mass of $\mu = m_\pi = 0.14 \,GeV$
 we obtain $m_{\rm  eff}\,=0.2\,GeV$. 

  %%%%%%%%%%%%%%%%%%%%%%%%%%%%%%%%%%%%%%%%%%%%%%%%%%%%%%%%%%%%%%%%%%%%%%%
%%%%%%%%%%%%%%%%%%%%%%%%%%%%%%%%%%%%%%%%%%%%%%%% %%%%%%%%%%%%%%%%%%%%

Discussing  hadron production we have to  construct a model for 
the 
 process of hadronization.  Our model  is the production of the gluon
 jets with the hadronization,  which is given by the fragmentation
 functions. We showed that in this model for  confinement, we
 obtained a reasonable description of the experimental data, with the
 thermal radiation and with the temperature of \eq{THTEM}, which
 is predicted in the CGC approach. 
It is possible that our hadronization model is too primitive, and
 for the gluon  with the transverse momenta of the order of
 $\Lambda_{\rm QCD}$, we should not apply the CGC formulae which
 are based on the  perturbative QCD approach. If  we  cut our gluon
 spectra at $p_T =\Lambda_{\rm QCD}$, we obtain a  good description
 of the experimental data, without the thermal radiation.  An
 alternate picture could be the following: The propagator of the
 gluon with transverse momentum $p_T$  in the  CGC medium with
 the temperature $T_{\rm th}$,  acquires a mass
 $m_g\,\,\propto\,\, T_{\rm th}$\cite{BRPI} and 
the propagator  acquires the form $1/(p^2_T + m^2_g)$.
 This mass provides the infrared cutoff in the gluon
 spectrum. Therefore, the same
rescatterings in the produced CGC medium, which generates
 the thermal spectrum can be a reason for blocking the small
 gluon $p_T \approx T_{\rm th} \approx 0.12-0.14 \,GeV$. In
 this picture we will not see any thermal emission in the
 spectrum of hadrons.  Note, that  $p_T \sim 0.12- 0.14\, GeV $
 corresponds to the $m_{\rm eff}  \approx 0.06\,GeV$ and to the
 distribution of \fig{meff}-b.

\begin{figure}[ht]
\begin{tabular}{c c }
 \includegraphics[width=7cm]{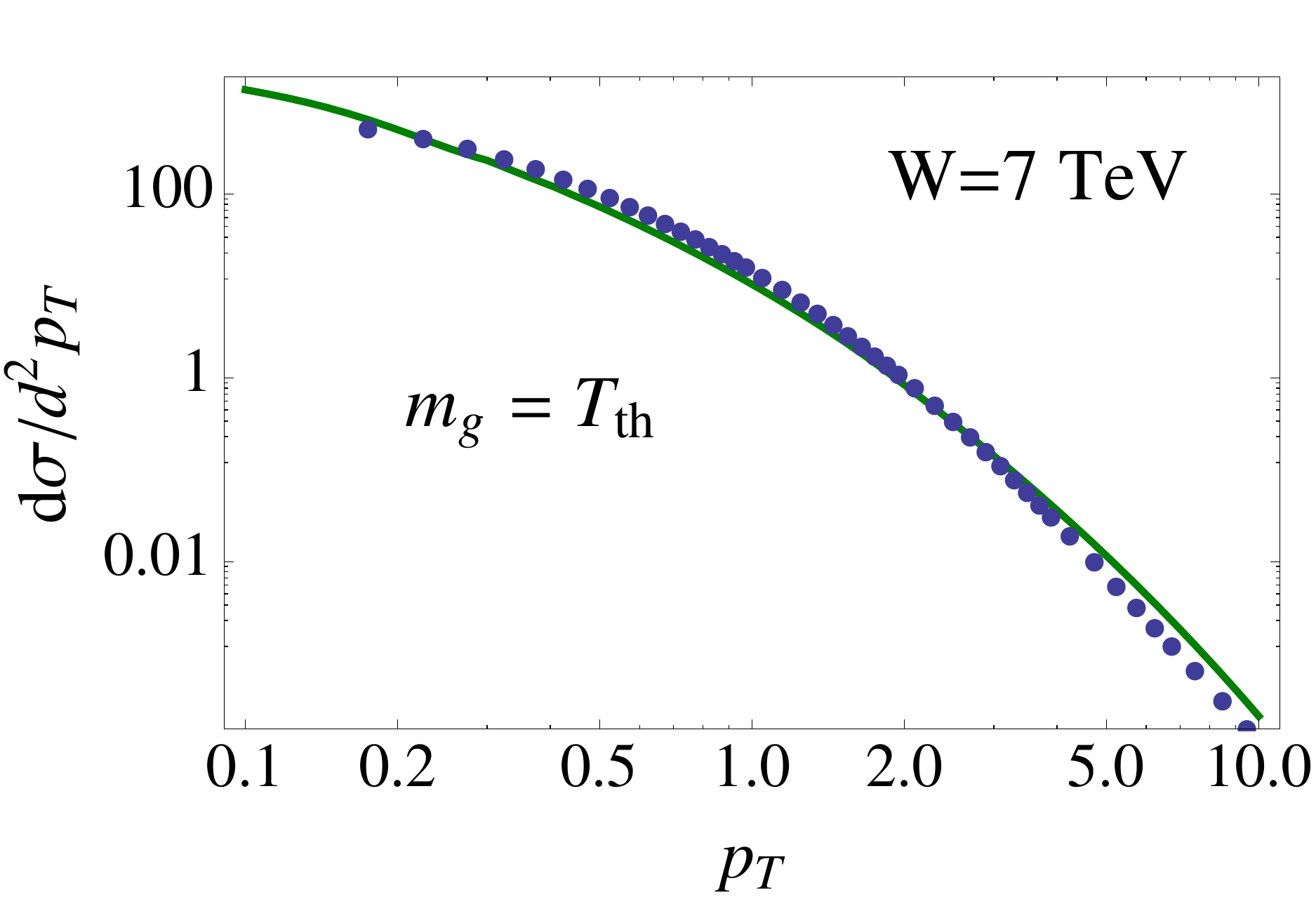}& \includegraphics[width=7cm]{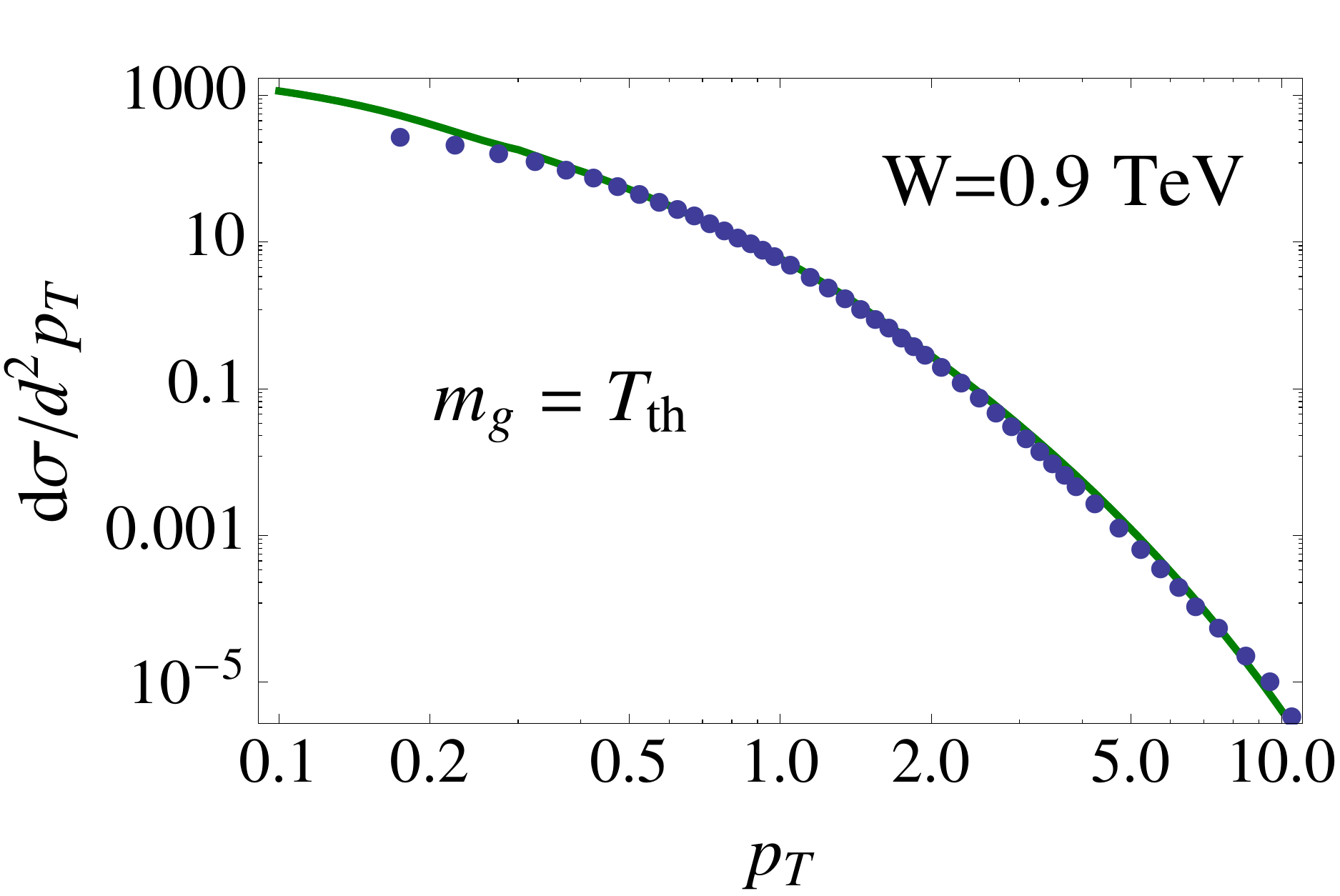}\\
 \fig{fitm}-a &  \fig{fitm}-b\\
 \end{tabular}
    \protect\caption{ $d^2 \sigma/d^2 p_T$ for W= 7 and 0.9 TeV with
 the gluon propagator $1/(p^2_T + m^2_g)$ and with $m_g =T_{\rm th}$.}
\label{fitm}
   \end{figure}

 In \fig{fitm} we present  the estimates  with the gluon
 propagator $1/(p^2_T + m^2)$ with $m = T_{\rm th}$. One
 can see that we are able to  successfully describe the data without
 the thermal radiation term.  Such a description should 
only be
 considered  with a grain of salt, since  $ m = g T$ with small 
$g$ in
 Ref.\cite{BRPI}, and realistic estimates will overshoot the data.

  %%%%%%%%%%%%%%%%%%%%%%%%%%%%%%%%%%%%%%%%%%%%%%%%%%%%%%%%%%%%%%%%%%%%%%%
%%%%%%%%%%%%%%%%%%%%%%%%%%%%%%%%%%%%%%%%%%%%%%%% %%%%%%%%%%%%%%%%%%%% 
 
\begin{figure}[ht]
\begin{tabular}{c c }
 \includegraphics[width=7cm]{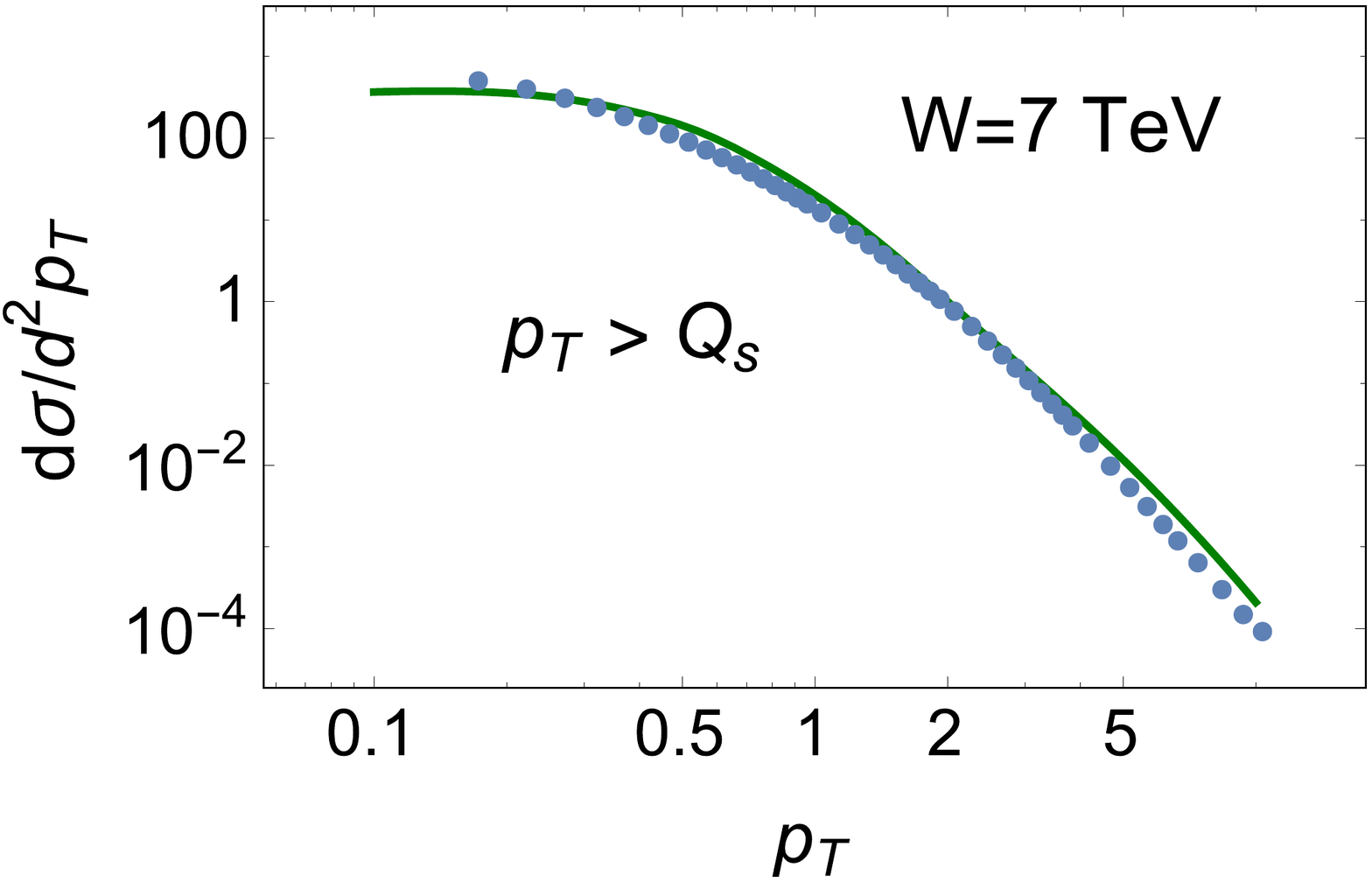}& \includegraphics[width=7cm]{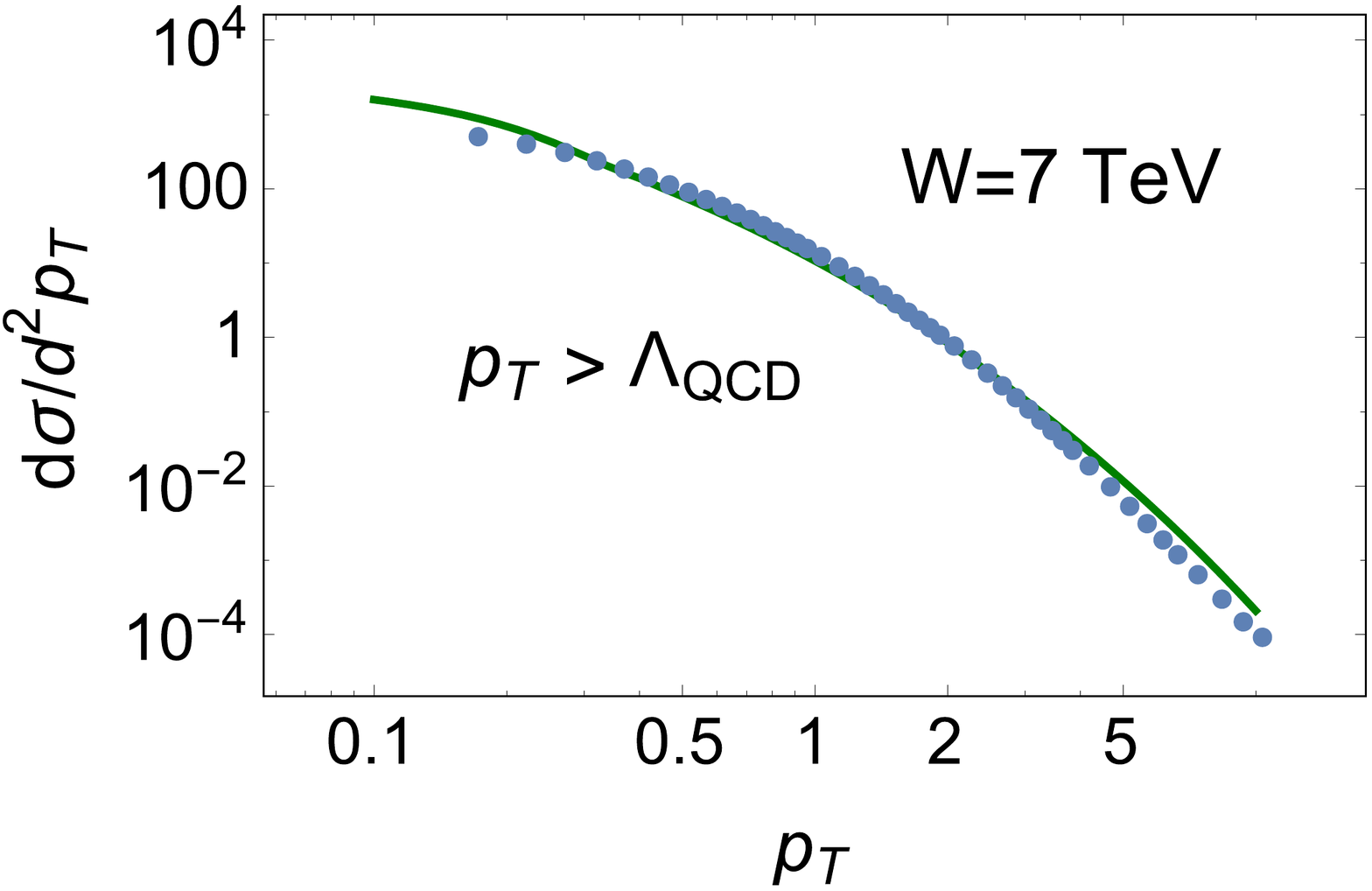}\\
 \fig{fitpl}-a &  \fig{fitpl}-b\\
 \end{tabular}
    \protect\caption{ $d^2 \sigma/d^2 p_T$ for W = 7  TeV  with the
 restrictions of $p_T$ of the  gluon jets:
    \fig{fitpl}-a only jets with $p_T > Q_s$  produce hadrons, and 
\fig{fitpl}-b gluons with $p_T < \Lambda_{QCD} = 200 \,MeV$ does not
 contribute to the production of hadrons.
 .}
\label{fitpl}
   \end{figure}

 In all the examples above,  we change our model for the 
hadronization,
 adding  the mass of the gluon jet. To illustrate our claim, that the
 existence of the thermal radiation crucially depends on the model of
  confinement, we plot in \fig{fitpl} the inclusive spectra for two
 different models: in \fig{fitpl}-a we assume that the gluons with
 $p_T < Q_s$ do not take part in the hadronization, and in
 \fig{fitpl}-b only gluons with $p_T > \Lambda_{QCD}$ produce the jet of
 hadrons.  \fig{fitpl}-a shows that we need the thermal radiation term
 with the contribution of 56\% to describe the data, while in 
\fig{fitpl}-b
 the data do not require the thermal emission.
 
   %%%%%%%%%%%%%%%%%%%%%%%%%%%%%%%%%%%%%%%%%%%%%%%%%%%%%%%%%%%%%%%%%%%%%%%
%%%%%%%%%%%%%%%%%%%%%%%%%%%%%%%%%%%%%%%%%%%%%%%% %%%%%%%%%%%%%%%%%%%% 

%%%%%%%%%%%%%%%%%%%%%%%%%%%%%%%%%%%%%%%%%%%%%%%%%%%%%% 
  \section{Conclusions}
  %%%%%%%%%%%%%%%%%%%%%%%%%%%%%%%%%%%%%%%%%%%%%%%%%%%%
  The main result of the paper is, that we show the need for  thermal
 emission within a particular model for  confinement: the parton
 (quark or gluon)  with the transverse momenta of the order of $Q_s$ 
decays into  hadrons with  the given fragmentation functions. The 
temperature of this emission turns out to be equal to 
  $2.3/(2 \,\pi)\,Q_s$, as  was expected in the  CGC/saturation
 approach. Note, that the coefficient $c$ in \eq{THTEM} turns out
 to be in almost two times larger than predicted in Ref.\cite{KLT}.

  We develop the formalism  for the calculation of the transverse
 momenta spectra in CGC/saturation approach, which is based on the
 observation that even for small values of $p_T$ the main contribution
 stems from the kinematic region in vicinity of the saturation momentum, 
where theoretically, we know the scattering amplitude.  In other words,
 it means that we do not need to 
introduce the non-perturbative corrections due to the unknown physics at long
 distances (see Refs.\cite{VAZW,KHALE} for example) in the dipole scattering
 amplitude. The non-perturbative corrections have to be included to 
describe
 the process of hadronization, which we discuss in the 
 model.  This model incorporates  the decay of the gluon jet with the
 effective mass  $m^2_{\rm eff} = 2 Q_s \mu_{\rm soft}$ where $\mu_{\rm
 soft}$ is the soft scale, and with the fragmentation functions of \eq{FRF}
 at all  values of the transverse momenta.

 We 
suggest to take
 into account the behaviour of $\nabla^2 N$ in the saturation region in
 the form:
  \beq \label{CON1}
  \nabla^2N\,\,=\,\,\nabla^2\Bigg( 1 - \exp\Lb - \phi_0 \Lb r^2\,Q^2_s\Rb^{\bar{\gamma}}\Rb\Bigg)
  \eeq
  and demonstrate that this suggestion follows from the solution to the
 non-linear Baltsky-Kovchegov equation, for the simplified BFKL kernel.

 It should be emphasized that we reproduce the experimental data without
 any K-factor, which is used  for accounting for 
 higher order
 corrections.  We wish also to  mention, that we have calculated the
 inclusive production taking $\bas = 0.25$. This value is less that
 $\bas\Lb Q_s\Rb\,\,=\,\,0.3$ which appears more natural in \eq{MF1}.
 For $\bas = \bas\Lb Q_s\Rb$, we need to introduce a K-factor of about
 1.3 - 1.5.

 The value of the thermal radiation term contribution
 depends on the value of the effective mass $m_{\rm eff}$, however, in the
 region of possible values for this mass $0.12 - 0.5 \,GeV$ we need
 to account for the thermal emission to describe the spectrum at low $p_T$.

   We show that a different mechanism of confinement that blocks
 the emission of gluons with  $p_T \,\leq \Lambda_{\rm QCD} $ or/and 
 that generates the gluon mass $m = T_{\rm th}$,  is able to describe
 the experimental data without the thermal radiation.
 
  Hence, we state that the existence of the thermal term in 
the $p_T$ spectrum of produced hadrons, depends crucially on the model
 for hadronization.
    
\begin{figure}[ht]
    \centering
  \leavevmode
      \includegraphics[width=8cm]{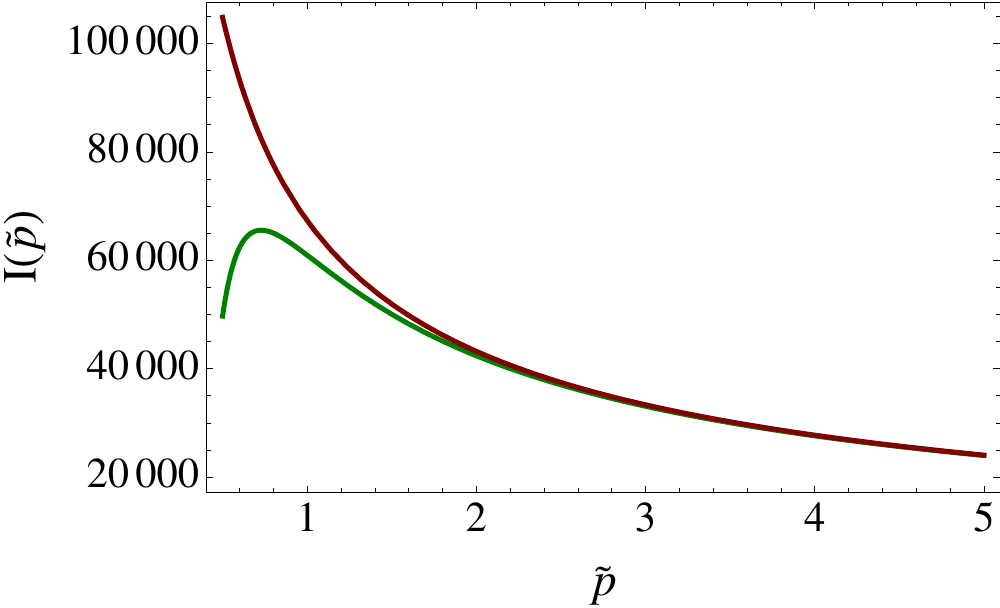}  
    \caption{ The green line is \eq{PP} while the red one describes
 only the first term of this equation..}
\label{strf}
  \end{figure}

  In our approach we are able to evaluate the  kinematic region that we 
can
 use the formalism of the deep inelastic structure functions. The structure
 function is related to the scattering amplitude being
 $\int d^2 b N\Lb Y, r,b\Rb$. However, as we have discussed
 the inclusive production is determined by function $N_G$ 
 (see \eq{NG}). In the region where we can neglect the
 $N^2$ -term in $N_G$, we can safely perform the integration
 over $b$, and obtain the expression for the inclusive production
 through the structure functions. In \fig{strf} we show \eq{PP}
 and the first term of this equation which is the  gluon structure
 function in the vicinity of the saturation scale. One can conclude
 that for $p_T \,>\,2 \,Q_s$ we can safely use the gluon structure
 function which has been measured for DIS at HERA. On the other hand,
 the thermal emission  comes from the region $p_T \,<\,2 \,Q_s$  and,
 therefore, the existence of this phenomenon depends completely  on 
the
 inclusive prodiction in CGC/saturation approach in this region. 
  
    %%%%%%%%%%%%%%%%%%%%%%%%%%%%%%%%%%%%%%%%%%%%%%%%%%%%%

 %%%%%%%%%%%%%%%%%%%%%%%%%%%%%%%%%%%%%%%%%%%%%%%%%%%%%%%%%%  
  %%%%%%%%%% %%%%%%%%%%%%%%%%%%%%%%%%%%%%%%%%%%%%%%%%%%%%%%%%%%%%%% 
  \section{Acknowledgements}
  %%%%%%%%%%%%%%%%%%%%%%%%%%%%%%%%%%%%%%%%%%%%%%%%%%%%%%%%%%%%%%% 
   We thank our colleagues at Tel Aviv university and UTFSM for
 encouraging discussions. Our special thanks go to  
  Keith Baker and Dmitry Kharzeev    for  fruitful 
 discussions on the subject which   prompted the appearance of this 
paper. 
 
  This research was supported  by 
   Proyecto Basal FB 0821(Chile),  Fondecyt (Chile) grant  
 1180118 and by   CONICYT grant PIA ACT1406.

\end{document}